

%
%
%
\def\unredoffs{} \def\redoffs{\voffset=-.31truein\hoffset=-.59truein}
\def\speclscape{\special{ps: landscape}}
%
%
%
%
\newbox\leftpage \newdimen\fullhsize \newdimen\hstitle \newdimen\hsbody
\tolerance=1000\hfuzz=2pt
\catcode`\@=11 
\def\bigans{b }
\def\answ{b }

%
\ifx\answ\bigans\message{(This will come out unreduced.}
\magnification=1200\unredoffs\baselineskip=16pt plus 2pt minus 1pt
\hsbody=\hsize \hstitle=\hsize 
\else\message{(This will be reduced.} \let\l@r=L
\magnification=1000\baselineskip=16pt plus 2pt minus 1pt \vsize=7truein
\redoffs \hstitle=8truein\hsbody=4.75truein\fullhsize=10truein\hsize=\hsbody
\output={\ifnum\pageno=0 
  \shipout\vbox{\speclscape{\hsize\fullhsize\makeheadline}
    \hbox to \fullhsize{\hfill\pagebody\hfill}}\advancepageno
  \else
  \almostshipout{\leftline{\vbox{\pagebody\makefootline}}}\advancepageno
  \fi}
\def\almostshipout#1{\if L\l@r \count1=1 \message{[\the\count0.\the\count1]}
      \global\setbox\leftpage=#1 \global\let\l@r=R
 \else \count1=2
  \shipout\vbox{\speclscape{\hsize\fullhsize\makeheadline}
      \hbox to\fullhsize{\box\leftpage\hfil#1}}  \global\let\l@r=L\fi}
\fi
%
\newcount\yearltd\yearltd=\year\advance\yearltd by -1900

\def\Title#1#2{\nopagenumbers\abstractfont\hsize=\hstitle\rightline{#1}%
\vskip 1in\centerline{\titlefont #2}\abstractfont\vskip .5in\pageno=0}
\def\Date#1{\vfill\leftline{#1}\tenpoint\supereject\global\hsize=\hsbody%
\footline={\hss\tenrm\folio\hss}}
%

\def\draftmode{\message{ DRAFTMODE }\def\draftdate{{\rm preliminary draft:
\number\month/\number\day/\number\yearltd\ \ \hourmin}}%
\headline={\hfil\draftdate}\writelabels\baselineskip=20pt plus 2pt minus 2pt
 {\count255=\time\divide\count255 by 60 \xdef\hourmin{\number\count255}
  \multiply\count255 by-60\advance\count255 by\time
  \xdef\hourmin{\hourmin:\ifnum\count255<10 0\fi\the\count255}}}
\def\nolabels{\def\wrlabeL##1{}\def\eqlabeL##1{}\def\reflabeL##1{}}
\def\writelabels{\def\wrlabeL##1{\leavevmode\vadjust{\rlap{\smash%
{\line{{\escapechar=` \hfill\rlap{\sevenrm\hskip.03in\string##1}}}}}}}%
\def\eqlabeL##1{{\escapechar-1\rlap{\sevenrm\hskip.05in\string##1}}}%
\def\reflabeL##1{\noexpand\llap{\noexpand\sevenrm\string\string\string##1}}}
\nolabels
%
\global\newcount\secno \global\secno=0
\global\newcount\meqno \global\meqno=1
\def\newsec#1{\global\advance\secno by1\message{(\the\secno. #1)}
\global\subsecno=0\eqnres@t\noindent{\bf\the\secno. #1}
\writetoca{{\secsym} {#1}}\par\nobreak\medskip\nobreak}
\def\eqnres@t{\xdef\secsym{\the\secno.}\global\meqno=1\bigbreak\bigskip}
\def\sequentialequations{\def\eqnres@t{\bigbreak}}\xdef\secsym{}
\global\newcount\subsecno \global\subsecno=0
\def\subsec#1{\global\advance\subsecno by1\message{(\secsym\the\subsecno. #1)}
\ifnum\lastpenalty>9000\else\bigbreak\fi
\noindent{\it\secsym\the\subsecno. #1}\writetoca{\string\quad
{\secsym\the\subsecno.} {#1}}\par\nobreak\medskip\nobreak}
\def\appendix#1#2{\global\meqno=1\global\subsecno=0\xdef\secsym{\hbox{#1.}}
\bigbreak\bigskip\noindent{\bf Appendix #1. #2}\message{(#1. #2)}
\writetoca{Appendix {#1.} {#2}}\par\nobreak\medskip\nobreak}
%
%
\def\eqnn#1{\xdef #1{(\secsym\the\meqno)}\writedef{#1\leftbracket#1}%
\global\advance\meqno by1\wrlabeL#1}
\def\eqna#1{\xdef #1##1{\hbox{$(\secsym\the\meqno##1)$}}
\writedef{#1\numbersign1\leftbracket#1{\numbersign1}}%
\global\advance\meqno by1\wrlabeL{#1$\{\}$}}
\def\eqn#1#2{\xdef #1{(\secsym\the\meqno)}\writedef{#1\leftbracket#1}%
\global\advance\meqno by1$$#2\eqno#1\eqlabeL#1$$}
%
\newskip\footskip\footskip14pt plus 1pt minus 1pt 
\def\footnotefont{\ninepoint}\def\f@t#1{\footnotefont #1\@foot}
\def\f@@t{\baselineskip\footskip\bgroup\footnotefont\aftergroup\@foot\let\next}
\setbox\strutbox=\hbox{\vrule height9.5pt depth4.5pt width0pt}
\global\newcount\ftno \global\ftno=0
\def\foot{\global\advance\ftno by1\footnote{$^{\the\ftno}$}}
%
\newwrite\ftfile
\def\footend{\def\foot{\global\advance\ftno by1\chardef\wfile=\ftfile
$^{\the\ftno}$\ifnum\ftno=1\immediate\openout\ftfile=foots.tmp\fi%
\immediate\write\ftfile{\noexpand\smallskip%
\noexpand\item{f\the\ftno:\ }\pctsign}\findarg}%
\def\footatend{\vfill\eject\immediate\closeout\ftfile{\parindent=20pt
\centerline{\bf Footnotes}\nobreak\bigskip\input foots.tmp }}}
\def\footatend{}
%
%
\global\newcount\refno \global\refno=1
\newwrite\rfile
\def\ref{[\the\refno]\nref}
\def\nref#1{\xdef#1{[\the\refno]}\writedef{#1\leftbracket#1}%
\ifnum\refno=1\immediate\openout\rfile=refs.tmp\fi
\global\advance\refno by1\chardef\wfile=\rfile\immediate
\write\rfile{\noexpand\item{#1\ }\reflabeL{#1\hskip.31in}\pctsign}\findarg}
\def\findarg#1#{\begingroup\obeylines\newlinechar=`\^^M\pass@rg}
{\obeylines\gdef\pass@rg#1{\writ@line\relax #1^^M\hbox{}^^M}%
\gdef\writ@line#1^^M{\expandafter\toks0\expandafter{\striprel@x #1}%
\edef\next{\the\toks0}\ifx\next\em@rk\let\next=\endgroup\else\ifx\next\empty%
\else\immediate\write\wfile{\the\toks0}\fi\let\next=\writ@line\fi\next\relax}}
\def\striprel@x#1{} \def\em@rk{\hbox{}}
\def\lref{\begingroup\obeylines\lr@f}
\def\lr@f#1#2{\gdef#1{\ref#1{#2}}\endgroup\unskip}

\def\addref#1{\immediate\write\rfile{\noexpand\item{}#1}} 
\def\startrefs#1{\immediate\openout\rfile=refs.tmp\refno=#1}
\def\xref{\expandafter\xr@f}\def\xr@f[#1]{#1}
\def\refs#1{\count255=1[\r@fs #1{\hbox{}}]}
\def\r@fs#1{\ifx\und@fined#1\message{reflabel \string#1 is undefined.}%
\nref#1{need to supply reference \string#1.}\fi%
\vphantom{\hphantom{#1}}\edef\next{#1}\ifx\next\em@rk\def\next{}%
\else\ifx\next#1\ifodd\count255\relax\xref#1\count255=0\fi%
\else#1\count255=1\fi\let\next=\r@fs\fi\next}
%

%
\newwrite\ffile\global\newcount\figno \global\figno=1
\def\fig{fig.~\the\figno\nfig}
\def\nfig#1{\xdef#1{fig.~\the\figno}%
\writedef{#1\leftbracket fig.\noexpand~\the\figno}%
\ifnum\figno=1\immediate\openout\ffile=figs.tmp\fi\chardef\wfile=\ffile%
\immediate\write\ffile{\noexpand\medskip\noexpand\item{Fig.\ \the\figno. }
\reflabeL{#1\hskip.55in}\pctsign}\global\advance\figno by1\findarg}
\def\xfig{\expandafter\xf@g}\def\xf@g fig.\penalty\@M\ {}
\def\figs#1{figs.~\f@gs #1{\hbox{}}}
\def\f@gs#1{\edef\next{#1}\ifx\next\em@rk\def\next{}\else
\ifx\next#1\xfig #1\else#1\fi\let\next=\f@gs\fi\next}
\newwrite\lfile
{\escapechar-1\xdef\pctsign{\string\%}\xdef\leftbracket{\string\{}
\xdef\rightbracket{\string\}}\xdef\numbersign{\string\#}}

\def\writestop{\def\writestoppt{\immediate\write\lfile{\string\pageno%
\the\pageno\string\startrefs\leftbracket\the\refno\rightbracket%
\string\def\string\secsym\leftbracket\secsym\rightbracket%
\string\secno\the\secno\string\meqno\the\meqno}\immediate\closeout\lfile}}
\def\writestoppt{}\def\writedef#1{}
\def\seclab#1{\xdef #1{\the\secno}\writedef{#1\leftbracket#1}\wrlabeL{#1=#1}}
\def\subseclab#1{\xdef #1{\secsym\the\subsecno}%
\writedef{#1\leftbracket#1}\wrlabeL{#1=#1}}
\newwrite\tfile \def\writetoca#1{}
\def\leaderfill{\leaders\hbox to 1em{\hss.\hss}\hfill}
\def\writetoc{\immediate\openout\tfile=toc.tmp
   \def\writetoca##1{{\edef\next{\write\tfile{\noindent ##1
   \string\leaderfill {\noexpand\number\pageno} \par}}\next}}}

%
\catcode`\@=12 
%
\edef\tfontsize{\ifx\answ\bigans scaled\magstep3\else scaled\magstep4\fi}
\font\titlerm=cmr10 \tfontsize \font\titlerms=cmr7 \tfontsize
\font\titlermss=cmr5 \tfontsize \font\titlei=cmmi10 \tfontsize
\font\titleis=cmmi7 \tfontsize \font\titleiss=cmmi5 \tfontsize
\font\titlesy=cmsy10 \tfontsize \font\titlesys=cmsy7 \tfontsize
\font\titlesyss=cmsy5 \tfontsize \font\titleit=cmti10 \tfontsize
\skewchar\titlei='177 \skewchar\titleis='177 \skewchar\titleiss='177
\skewchar\titlesy='60 \skewchar\titlesys='60 \skewchar\titlesyss='60
\def\titlefont{\def\rm{\fam0\titlerm}
\textfont0=\titlerm \scriptfont0=\titlerms \scriptscriptfont0=\titlermss
\textfont1=\titlei \scriptfont1=\titleis \scriptscriptfont1=\titleiss
\textfont2=\titlesy \scriptfont2=\titlesys \scriptscriptfont2=\titlesyss
\textfont\itfam=\titleit \def\it{\fam\itfam\titleit}\rm}
 \ifx\answ\bigans\else scaled\magstep1\fi
\ifx\answ\bigans\def\abstractfont{\tenpoint}\else
\font\abssl=cmsl10 scaled \magstep1
\font\absrm=cmr10 scaled\magstep1 \font\absrms=cmr7 scaled\magstep1
\font\absrmss=cmr5 scaled\magstep1 \font\absi=cmmi10 scaled\magstep1
\font\absis=cmmi7 scaled\magstep1 \font\absiss=cmmi5 scaled\magstep1
\font\abssy=cmsy10 scaled\magstep1 \font\abssys=cmsy7 scaled\magstep1
\font\abssyss=cmsy5 scaled\magstep1 \font\absbf=cmbx10 scaled\magstep1
\skewchar\absi='177 \skewchar\absis='177 \skewchar\absiss='177
\skewchar\abssy='60 \skewchar\abssys='60 \skewchar\abssyss='60
\def\abstractfont{\def\rm{\fam0\absrm}
\textfont0=\absrm \scriptfont0=\absrms \scriptscriptfont0=\absrmss
\textfont1=\absi \scriptfont1=\absis \scriptscriptfont1=\absiss
\textfont2=\abssy \scriptfont2=\abssys \scriptscriptfont2=\abssyss
\textfont\itfam=\bigit \def\it{\fam\itfam\bigit}\def\footnotefont{\tenpoint}%
\textfont\slfam=\abssl \def\sl{\fam\slfam\abssl}%
\textfont\bffam=\absbf \def\bf{\fam\bffam\absbf}\rm}\fi
\def\tenpoint{\def\rm{\fam0\tenrm}
\textfont0=\tenrm \scriptfont0=\sevenrm \scriptscriptfont0=\fiverm
\textfont1=\teni  \scriptfont1=\seveni  \scriptscriptfont1=\fivei
\textfont2=\tensy \scriptfont2=\sevensy \scriptscriptfont2=\fivesy
\textfont\itfam=\tenit \def\it{\fam\itfam\tenit}\def\footnotefont{\ninepoint}%
\textfont\bffam=\tenbf \def\bf{\fam\bffam\tenbf}\def\sl{\fam\slfam\tensl}\rm}
\font\ninerm=cmr9 \font\sixrm=cmr6 \font\ninei=cmmi9 \font\sixi=cmmi6
\font\ninesy=cmsy9 \font\sixsy=cmsy6 \font\ninebf=cmbx9
\font\nineit=cmti9 \font\ninesl=cmsl9 \skewchar\ninei='177
\skewchar\sixi='177 \skewchar\ninesy='60 \skewchar\sixsy='60
\def\ninepoint{\def\rm{\fam0\ninerm}
\textfont0=\ninerm \scriptfont0=\sixrm \scriptscriptfont0=\fiverm
\textfont1=\ninei \scriptfont1=\sixi \scriptscriptfont1=\fivei
\textfont2=\ninesy \scriptfont2=\sixsy \scriptscriptfont2=\fivesy
\textfont\itfam=\ninei \def\it{\fam\itfam\nineit}\def\sl{\fam\slfam\ninesl}%
\textfont\bffam=\ninebf \def\bf{\fam\bffam\ninebf}\rm}
%
%

\hyphenation{anom-aly anom-alies coun-ter-term coun-ter-terms}
\def\inv{^{\raise.15ex\hbox{${\scriptscriptstyle -}$}\kern-.05em 1}}

\def\Dsl{\,\raise.15ex\hbox{/}\mkern-13.5mu D} 
\def\dsl{\raise.15ex\hbox{/}\kern-.57em\partial}

\def\tr{{\rm tr}} \def\Tr{{\rm Tr}}
\font\bigit=cmti10 scaled \magstep1
\def\lspace{\ifx\answ\bigans{}\else\qquad\fi}
\def\lbspace{\ifx\answ\bigans{}\else\hskip-.2in\fi} 
\def\boxeqn#1{\vcenter{\vbox{\hrule\hbox{\vrule\kern3pt\vbox{\kern3pt
    \hbox{${\displaystyle #1}$}\kern3pt}\kern3pt\vrule}\hrule}}}
\def\mbox#1#2{\vcenter{\hrule \hbox{\vrule height#2in
        \kern#1in \vrule} \hrule}}  
%

\def\darr#1{\raise1.5ex\hbox{$\leftrightarrow$}\mkern-16.5mu #1}

\def\roughly#1{\raise.3ex\hbox{$#1$\kern-.75em\lower1ex\hbox{$\sim$}}}

\let\includefigures=\iftrue
\let\useblackboard=\iftrue
\newfam\black

\includefigures
\message{If you do not have epsf.tex (to include figures),}
\message{change the option at the top of the tex file.}
\input epsf
\def\figin{\epsfcheck\figin}\def\figins{\epsfcheck\figins}
\def\epsfcheck{\ifx\epsfbox\UnDeFiNeD
\message{(NO epsf.tex, FIGURES WILL BE IGNORED)}
\gdef\figin##1{\vskip2in}\gdef\figins##1{\hskip.5in}
\else\message{(FIGURES WILL BE INCLUDED)}%
\gdef\figin##1{##1}\gdef\figins##1{##1}\fi}
\def\DefWarn#1{}
\def\figinsert{\goodbreak\midinsert}
\def\ifig#1#2#3{\DefWarn#1\xdef#1{fig.~\the\figno}
\writedef{#1\leftbracket fig.\noexpand~\the\figno}%
\figinsert\figin{\centerline{#3}}\medskip\centerline{\vbox{
\baselineskip12pt\advance\hsize by -1truein
\noindent\footnotefont{\bf Fig.~\the\figno:} #2}}
\endinsert\global\advance\figno by1}
\else
\def\ifig#1#2#3{\xdef#1{fig.~\the\figno}
\writedef{#1\leftbracket fig.\noexpand~\the\figno}%
\global\advance\figno by1} \fi

\def\id{{1 \kern-.28em {\rm l}}}

\def\CM{{\cal M}}

\def\K3{{\bf K3}}
\def\journal#1&#2(#3){\unskip, \sl #1\ \bf #2 \rm(19#3) }
\def\andjournal#1&#2(#3){\sl #1~\bf #2 \rm (19#3) }

\def\bar{\overline}
\def\hat{\widehat}
\def\ie{{\it i.e.}}
\def\eg{{\it e.g.}}

\def\tilde{\widetilde}

\def\frac#1#2{{#1\over#2}}

\def\inbar{\,\vrule height1.5ex width.4pt depth0pt}
\def\IC{\relax\hbox{$\inbar\kern-.3em{\rm C}$}}
\def\IR{\relax{\rm I\kern-.18em R}}
\def\IP{\relax{\rm I\kern-.18em P}}

%
%

%
\catcode`\@=11
\def\slash#1{\mathord{\mathpalette\c@ncel{#1}}}
\overfullrule=0pt

\def\FF{{\cal F}}
\def\GG{{\cal G}}

\def\NN{{\cal N}}
\def\OO{{\cal O}}

\def\SS{{\cal S}}

\def\underrel#1\over#2{\mathrel{\mathop{\kern\z@#1}\limits_{#2}}}

\catcode`\@=12


%

\def\det{{\rm det}}
\def\tr{{\rm tr}}
\def\Tr{{\rm Tr}}

\def\det{{\rm det}}
\def\exp{{\rm exp}}


\def\p{{\partial}}

\def\ft{{\tilde f}}

\def\rt{{\tilde{r}}}

\lref\SakaiGF{
  T.~Sakai and S.~Sugimoto,
  ``Low energy hadron physics in holographic QCD,''
}
\lref\ErlichQH{
  J.~Erlich, E.~Katz, D.~T.~Son and M.~A.~Stephanov,
  ``QCD and a holographic model of hadrons,''
Phys.\ Rev.\ Lett.\  {\bf 95}, 261602 (2005). [hep-ph/0501128].
}
\lref\JarvinenQE{
  M.~Jarvinen and E.~Kiritsis,
  ``Holographic Models for QCD in the Veneziano Limit,''
JHEP {\bf 1203}, 002 (2012). [arXiv:1112.1261 [hep-ph]].
}
\lref\CaseroAE{
  R.~Casero, E.~Kiritsis and A.~Paredes,
  ``Chiral symmetry breaking as open string tachyon condensation,''
Nucl.\ Phys.\ B {\bf 787}, 98 (2007). [hep-th/0702155 [HEP-TH]].
}

\lref\DymarskyUH{
  A.~Dymarsky, I.~R.~Klebanov and R.~Roiban,
  ``Perturbative search for fixed lines in large N gauge theories,''
JHEP {\bf 0508}, 011 (2005).
[hep-th/0505099].
}

\lref\DymarskyNC{
  A.~Dymarsky, I.~R.~Klebanov and R.~Roiban,
  ``Perturbative gauge theory and closed string tachyons,''
JHEP {\bf 0511}, 038 (2005).
[hep-th/0509132].
}

\lref\PomoniDE{
  E.~Pomoni and L.~Rastelli,
  ``Large N Field Theory and AdS Tachyons,''
JHEP {\bf 0904}, 020 (2009). [arXiv:0805.2261 [hep-th]].
}

\lref\MiranskyPD{
  V.~A.~Miransky and K.~Yamawaki,
  ``Conformal phase transition in gauge theories,''
Phys.\ Rev.\ D {\bf 55}, 5051 (1997), [Erratum-ibid.\ D {\bf 56},
3768 (1997)]. [hep-th/9611142].
}
\lref\KachruYS{
  S.~Kachru and E.~Silverstein,
  ``4-D conformal theories and strings on orbifolds,''
Phys.\ Rev.\ Lett.\  {\bf 80}, 4855 (1998). [hep-th/9802183].
}
\lref\LawrenceJA{
  A.~E.~Lawrence, N.~Nekrasov and C.~Vafa,
  ``On conformal field theories in four-dimensions,''
Nucl.\ Phys.\ B {\bf 533}, 199 (1998). [hep-th/9803015].
}
\lref\LuninJY{
  O.~Lunin and J.~M.~Maldacena,
  ``Deforming field theories with U(1) x U(1) global symmetry and their gravity duals,''
JHEP {\bf 0505}, 033 (2005). [hep-th/0502086].
}
\lref\FaulknerJY{
  T.~Faulkner, H.~Liu and M.~Rangamani,
  ``Integrating out geometry: Holographic Wilsonian RG and the membrane paradigm,''
JHEP {\bf 1108}, 051 (2011). [arXiv:1010.4036 [hep-th]].
}
\lref\FreedmanGP{
  D.~Z.~Freedman, S.~S.~Gubser, K.~Pilch and N.~P.~Warner,
 ``Renormalization group flows from holography supersymmetry and a c theorem,''
Adv.\ Theor.\ Math.\ Phys.\  {\bf 3}, 363 (1999). [hep-th/9904017].
}
\lref\deBoerXF{
  J.~de Boer, E.~P.~Verlinde and H.~L.~Verlinde,
  ``On the holographic renormalization group,''
JHEP {\bf 0008}, 003 (2000). [hep-th/9912012].
}
\lref\deBoerCZ{
  J.~de Boer,
  ``The Holographic renormalization group,''
Fortsch.\ Phys.\  {\bf 49}, 339 (2001). [hep-th/0101026].
}
\lref\HeemskerkHK{
  I.~Heemskerk and J.~Polchinski,
  ``Holographic and Wilsonian Renormalization Groups,''
JHEP {\bf 1106}, 031 (2011). [arXiv:1010.1264 [hep-th]].
}
\lref\NickelPR{
  D.~Nickel and D.~T.~Son,
  ``Deconstructing holographic liquids,''
New J.\ Phys.\  {\bf 13}, 075010 (2011). [arXiv:1009.3094 [hep-th]].
}
\lref\CohenSQ{
  A.~G.~Cohen and H.~Georgi,
  ``Walking Beyond The Rainbow,''
Nucl.\ Phys.\ B {\bf 314}, 7 (1989)..
}
\lref\KutasovUQ{
  D.~Kutasov, J.~Lin and A.~Parnachev,
  ``Holographic Walking from Tachyon DBI,''
Nucl.\ Phys.\ B {\bf 863}, 361 (2012). [arXiv:1201.4123 [hep-th]].
}
\lref\VenezianoWM{
  G.~Veneziano,
  ``Some Aspects of a Unified Approach to Gauge, Dual and Gribov Theories,''
Nucl.\ Phys.\ B {\bf 117}, 519 (1976)..
}
\lref\VenezianoEC{
  G.~Veneziano,
  ``U(1) Without Instantons,''
Nucl.\ Phys.\ B {\bf 159}, 213 (1979)..
}
\lref\KosterlitzSM{
  J.~M.~Kosterlitz,
  ``The Critical properties of the two-dimensional x y model,''
J.\ Phys.\ C {\bf 7}, 1046 (1974).. }
\lref\KutasovFR{
  D.~Kutasov, J.~Lin and A.~Parnachev,
  ``Conformal Phase Transitions at Weak and Strong Coupling,''
Nucl.\ Phys.\ B {\bf 858}, 155 (2012). [arXiv:1107.2324 [hep-th]].
}
\lref\KaplanKR{
  D.~B.~Kaplan, J.~-W.~Lee, D.~T.~Son and M.~A.~Stephanov,
  ``Conformality Lost,''
Phys.\ Rev.\ D {\bf 80}, 125005 (2009). [arXiv:0905.4752 [hep-th]].
}
\lref\JensenGA{
  K.~Jensen, A.~Karch, D.~T.~Son and E.~G.~Thompson,
  ``Holographic Berezinskii-Kosterlitz-Thouless Transitions,''
Phys.\ Rev.\ Lett.\  {\bf 105}, 041601 (2010). [arXiv:1002.3159
[hep-th]].
}
\lref\JensenVX{
  K.~Jensen,
  ``More Holographic Berezinskii-Kosterlitz-Thouless Transitions,''
Phys.\ Rev.\ D {\bf 82}, 046005 (2010). [arXiv:1006.3066 [hep-th]].
}
\lref\IqbalAJ{
  N.~Iqbal, H.~Liu and M.~Mezei,
  ``Quantum phase transitions in semi-local quantum liquids,''
[arXiv:1108.0425 [hep-th]].
}
\lref\WittenUA{
  E.~Witten,
  ``Multitrace operators, boundary conditions, and AdS / CFT correspondence,''
[hep-th/0112258].
}

\lref\EvansAT{
  N.~Evans,
  ``Holographic Description of the QCD Phase Diagram and Out of Equilibrium Dynamics,''
[arXiv:1209.0626 [hep-ph]].
}
\lref\ColemanMX{
  S.~R.~Coleman and E.~Witten,
  ``Chiral Symmetry Breakdown in Large N Chromodynamics,''
Phys.\ Rev.\ Lett.\  {\bf 45}, 100 (1980)..
}
\lref\Gianotti{
  F. Gianotti, CERN Seminar, "Update on the Standard Model Higgs searches in ATLAS",
July, 4 2012. ATLAS-CONF-2012-093 }
%
\lref\Incandela{
  J. Incandela, CERN Seminar, "Update on the Standard Model Higgs searches in CMS",
July, 4 2012.}
%
\lref\PeskinZT{
  M.~E.~Peskin and T.~Takeuchi,
  ``A New constraint on a strongly interacting Higgs sector,''
Phys.\ Rev.\ Lett.\  {\bf 65}, 964 (1990)..
}
\lref\PeskinSW{
  M.~E.~Peskin and T.~Takeuchi,
  ``Estimation of oblique electroweak corrections,''
Phys.\ Rev.\ D {\bf 46}, 381 (1992)..
}
\lref\KennedySN{
  D.~C.~Kennedy and B.~W.~Lynn,
  ``Electroweak Radiative Corrections with an Effective Lagrangian: Four Fermion Processes,''
Nucl.\ Phys.\ B {\bf 322}, 1 (1989)..
}
\lref\CaroneMD{
  C.~D.~Carone, J.~Erlich and M.~Sher,
  ``Holographic Electroweak Symmetry Breaking from D-branes,''
Phys.\ Rev.\ D {\bf 76}, 015015 (2007). [arXiv:0704.3084 [hep-th]].
}
\lref\HirayamaHZ{
  T.~Hirayama and K.~Yoshioka,
  ``Holographic Construction of Technicolor Theory,''
JHEP {\bf 0710}, 002 (2007). [arXiv:0705.3533 [hep-ph]].
}
\lref\DietrichUP{
  D.~D.~Dietrich and C.~Kouvaris,
  ``Generalised bottom-up holography and walking technicolour,''
Phys.\ Rev.\ D {\bf 79}, 075004 (2009). [arXiv:0809.1324 [hep-ph]].
}
\lref\CaroneRX{
  C.~D.~Carone, J.~Erlich and M.~Sher,
  ``Extra Gauge Invariance from an Extra Dimension,''
Phys.\ Rev.\ D {\bf 78}, 015001 (2008). [arXiv:0802.3702 [hep-ph]].
}
\lref\SanninoHA{
  F.~Sannino,
  ``Dynamical Stabilization of the Fermi Scale: Phase Diagram of Strongly Coupled Theories for (Minimal) Walking Technicolor and Unparticles,''
[arXiv:0804.0182 [hep-ph]].
}
\lref\DietrichNI{
  D.~D.~Dietrich and C.~Kouvaris,
  ``Constraining vectors and axial-vectors in walking technicolour by a holographic principle,''
Phys.\ Rev.\ D {\bf 78}, 055005 (2008). [arXiv:0805.1503 [hep-ph]].
}
\lref\HabaNZ{
  K.~Haba, S.~Matsuzaki and K.~Yamawaki,
  ``S Parameter in the Holographic Walking/Conformal Technicolor,''
Prog.\ Theor.\ Phys.\  {\bf 120}, 691 (2008). [arXiv:0804.3668
[hep-ph]].
}
\lref\MintakevichWZ{
  O.~Mintakevich and J.~Sonnenschein,
  ``Holographic technicolor models and their S-parameter,''
JHEP {\bf 0907}, 032 (2009). [arXiv:0905.3284 [hep-th]].
}
\lref\FukanoZM{
  H.~S.~Fukano and F.~Sannino,
  ``Minimal Flavor Constraints for Technicolor,''
Int.\ J.\ Mod.\ Phys.\ A {\bf 25}, 3911 (2010). [arXiv:0908.2424
[hep-ph]].
}
\lref\KitazawaKR{
  N.~Kitazawa,
  ``Dynamical Electroweak Symmetry Breaking in String Models with D-branes,''
Int.\ J.\ Mod.\ Phys.\ A {\bf 25}, 2679 (2010). [arXiv:0908.2663
[hep-th]].
}
\lref\DietrichAF{
  D.~D.~Dietrich, M.~Jarvinen and C.~Kouvaris,
  ``Mixing in the axial sector in bottom-up holography for walking technicolour,''
JHEP {\bf 1007}, 023 (2010). [arXiv:0908.4357 [hep-ph]].
}
\lref\BelitskyFJ{
  A.~V.~Belitsky,
  ``Dual technicolor with hidden local symmetry,''
Phys.\ Rev.\ D {\bf 82}, 045006 (2010). [arXiv:1003.0062 [hep-ph]].
}
\lref\CaroneCP{
  C.~D.~Carone and R.~Primulando,
  ``Combined Constraints on Holographic Bosonic Technicolor,''
Phys.\ Rev.\ D {\bf 82}, 015003 (2010). [arXiv:1003.4720 [hep-ph]].
}
\lref\ReeceXJ{
  M.~Reece and L.~-T.~Wang,
  ``Randall-Sundrum and Strings,''
JHEP {\bf 1007}, 040 (2010). [arXiv:1003.5669 [hep-ph]].
}
\lref\AnguelovaQH{
  L.~Anguelova,
  ``Electroweak Symmetry Breaking from Gauge/Gravity Duality,''
Nucl.\ Phys.\ B {\bf 843}, 429 (2011). [arXiv:1006.3570 [hep-th]].
}
\lref\AnguelovaBC{
  L.~Anguelova, P.~Suranyi and L.~C.~R.~Wijewardhana,
  ``Holographic Walking Technicolor from D-branes,''
Nucl.\ Phys.\ B {\bf 852}, 39 (2011). [arXiv:1105.4185 [hep-th]].
}
\lref\LevkovYK{
  D.~G.~Levkov, V.~A.~Rubakov, S.~V.~Troitsky and Y.~A.~Zenkevich,
  ``Constraining holographic technicolor,''
[arXiv:1201.6368 [hep-ph]].
}
\lref\AnguelovaKA{
  L.~Anguelova, P.~Suranyi and L.~C.~R.~Wijewardhana,
  ``Scalar Mesons in Holographic Walking Technicolor,''
Nucl.\ Phys.\ B {\bf 862}, 671 (2012). [arXiv:1203.1968 [hep-th]].
}

%
\lref\LevkovAZ{
  D.~G.~Levkov, V.~A.~Rubakov, S.~V.~Troitsky and Y.~A.~Zenkevich,
  ``Constraining Holographic Technicolor,''
[arXiv:1201.6368 [hep-ph]].
}
\lref\HabaHU{
  K.~Haba, S.~Matsuzaki and K.~Yamawaki,
  ``Holographic Techni-dilaton,''
Phys.\ Rev.\ D {\bf 82}, 055007 (2010). [arXiv:1006.2526 [hep-ph]].
}
\lref\NunezWI{
  C.~Nunez, I.~Papadimitriou and M.~Piai,
  ``Walking Dynamics from String Duals,''
Int.\ J.\ Mod.\ Phys.\ A {\bf 25}, 2837 (2010). [arXiv:0812.3655
[hep-th]].
}

\lref\AgasheMC{
  K.~Agashe, C.~Csaki, C.~Grojean and M.~Reece,
  ``The S-parameter in holographic technicolor models,''
JHEP {\bf 0712}, 003 (2007). [arXiv:0704.1821 [hep-ph]].
}
\lref\CasalbuoniXN{
  R.~Casalbuoni, S.~De Curtis, D.~Dominici and D.~Dolce,
  ``Holographic approach to a minimal Higgsless model,''
JHEP {\bf 0708}, 053 (2007). [arXiv:0705.2510 [hep-ph]].
}
\lref\RoundNH{
  M.~Round,
  ``Generalised Holographic Electroweak Symmetry Breaking Models and the Possibility of Negative S,''
Phys.\ Rev.\ D {\bf 84}, 013012 (2011). [arXiv:1104.4037 [hep-ph]].
}
\lref\SondergaardPS{
  U.~I.~Sondergaard, C.~Pica and F.~Sannino,
  ``S-parameter at Non-Zero Temperature and Chemical Potential,''
Phys.\ Rev.\ D {\bf 84}, 075022 (2011). [arXiv:1107.1802 [hep-ph]].
}
\lref\CaroneCD{
  C.~D.~Carone,
  ``Technicolor with a 125 GeV Higgs Boson,''
[arXiv:1206.4324 [hep-ph]].
}
\lref\LawranceCG{
  R.~Lawrance and M.~Piai,
  ``Holographic Technidilaton and LHC searches,''
[arXiv:1207.0427 [hep-ph]].
}
\lref\MatsuzakiMK{
  S.~Matsuzaki and K.~Yamawaki,
  ``Is 125 GeV techni-dilaton found at LHC?,''
[arXiv:1207.5911 [hep-ph]].
}
\lref\ElanderFK{
  D.~Elander and M.~Piai,
  ``The decay constant of the holographic techni-dilaton and the 125 GeV boson,''
[arXiv:1208.0546 [hep-ph]].
}

\lref\HirnNT{
  J.~Hirn and V.~Sanz,
  ``A Negative S parameter from holographic technicolor,''
Phys.\ Rev.\ Lett.\  {\bf 97}, 121803 (2006).
[hep-ph/0606086].
}

\lref\PiaiHY{
  M.~Piai,
  ``Precision electro-weak parameters from AdS(5), localized kinetic terms and anomalous dimensions,''
[hep-ph/0608241].
}

\lref\CaroneWJ{
  C.~D.~Carone, J.~Erlich and J.~A.~Tan,
  ``Holographic Bosonic Technicolor,''
Phys.\ Rev.\ D {\bf 75}, 075005 (2007).
[hep-ph/0612242].
}

\lref\FabbrichesiGA{
  M.~Fabbrichesi, M.~Piai and L.~Vecchi,
  ``Dynamical electro-weak symmetry breaking from deformed AdS: Vector mesons and effective couplings,''
Phys.\ Rev.\ D {\bf 78}, 045009 (2008).
[arXiv:0804.0124 [hep-ph]].
}

\lref\PiaiMA{
  M.~Piai,
  ``Lectures on walking technicolor, holography and gauge/gravity dualities,''
Adv.\ High Energy Phys.\  {\bf 2010}, 464302 (2010).
[arXiv:1004.0176 [hep-ph]].
}

\lref\PeskinEV{
  M.~E.~Peskin and D.~V.~Schroeder,
  ``An Introduction to quantum field theory,''
Reading, USA: Addison-Wesley (1995) 842 p.
}

\lref\EvansJX{
  N.~Evans, J.~French and K.~-Y.~Kim,
  ``Inflation from Strongly Coupled Gauge Dynamics,''
[arXiv:1208.3060 [hep-th]].
}

\lref\AlvaresKR{
  R.~Alvares, N.~Evans and K.~-Y.~Kim,
  ``Holography of the Conformal Window,''
Phys.\ Rev.\ D {\bf 86}, 026008 (2012).
[arXiv:1204.2474 [hep-ph]].
}

\lref\RosensteinNM{
  B.~Rosenstein, B.~Warr and S.~H.~Park,
  ``Dynamical symmetry breaking in four Fermi interaction models,''
Phys.\ Rept.\  {\bf 205}, 59 (1991).
}

\lref\SSLee{
S.-S.~Lee,
"Low-energy effective theory of Fermi surface coupled with U(1) gauge field in 2+1 dimensions,"
Phys.\ Rev.\ B {\bf 80}, 165102 (2009).
[arXiv:0905.4532[cond-mat]].
}

\lref\StephanovWX{
  M.~A.~Stephanov,
  ``QCD phase diagram and the critical point,''
Prog.\ Theor.\ Phys.\ Suppl.\  {\bf 153}, 139 (2004), [Int.\ J.\ Mod.\ Phys.\ A {\bf 20}, 4387 (2005)].
[hep-ph/0402115].
}

\lref\CasalbuoniRS{
  R.~Casalbuoni,
  ``QCD critical point: A Historical perspective,''
PoS CPOD {\bf 2006}, 001 (2006).
[hep-ph/0610179].
}

\lref\HongSI{
  D.~K.~Hong and H.~-U.~Yee,
  ``Holographic estimate of oblique corrections for technicolor,''
Phys.\ Rev.\ D {\bf 74}, 015011 (2006).
[hep-ph/0602177].
}

\lref\AlhoMH{
  T.~Alho, M.~Jarvinen, K.~Kajantie, E.~Kiritsis and K.~Tuominen,
  ``On finite-temperature holographic QCD in the Veneziano limit,''
[arXiv:1210.4516 [hep-ph]].
}

\lref\MatsuzakiXX{
  S.~Matsuzaki and K.~Yamawaki,
  ``Holographic techni-dilaton at 125 GeV,''
[arXiv:1209.2017 [hep-ph]].
}

\lref\ChackoVM{
  Z.~Chacko, R.~Franceschini and R.~K.~Mishra,
  ``Resonance at 125 GeV: Higgs or Dilaton/Radion?,''
[arXiv:1209.3259 [hep-ph]].
}

\lref\BellazziniVZ{
  B.~Bellazzini, C.~Csaki, J.~Hubisz, J.~Serra and J.~Terning,
  ``A Higgslike Dilaton,''
[arXiv:1209.3299 [hep-ph]].
}


\Title{} {\vbox{\centerline{S-parameter, Technimesons, and Phase Transitions}
\bigskip
\centerline{in Holographic Tachyon DBI Models} }}
\bigskip

\centerline{\it Mikhail Goykhman  and Andrei Parnachev}
\bigskip
\smallskip
\centerline{Institute Lorentz for Theoretical Physics, Leiden
University} \centerline{P.O. Box 9506, Leiden 2300RA, The
Netherlands}
\smallskip

\vglue .3cm

\bigskip

\let\includefigures=\iftrue

\noindent We investigate some phenomenological aspects of the
holographic models based on the tachyon Dirac-Born-Infeld action in
the AdS space-time. These holographic  theories model strongly interacting fermions
and feature dynamical mass generation and symmetry breaking. We show
that they can be viewed as models of holographic walking technicolor
and compute the Peskin-Takeuchi S-parameter and  masses of lightest technimesons for a variety of the
tachyon potentials. We also investigate the phase structure at
finite temperature and charge density. Finally, we comment on the
holographic Wilsonian RG in the context of holographic tachyon DBI
models.

\bigskip

\Date{November, 2012}

\newsec{Introduction and summary}

\noindent Systems of strongly interacting fermions have applications in many realms, including
condensed matter (e.g. graphene) and particle physics (e.g. technicolor models).
A  simple way to introduce interaction between fermions involves adding a quartic term to the
lagrangian of $N$ free fermions, resulting in  the Nambu-Jona-Lasinio model (see e.g. \RosensteinNM\ for a review).
In three space-time dimensions the model is renormalizable to all orders in the $1/N$ expansion:
one can take a double scaling limit where  the coupling is tuned to the critical value, while the
UV cutoff is sent to infinity, keeping the physical mass fixed.
Dynamical mass generation at sufficiently large values of the coupling is an important feature
which is believed to happen in other strongly interacting fermion systems.

 Unfortunately one often has to resort to approximate methods to describe the physics in the
 vicinity of the phase transition from the massless phase to the one with a gap. This is because the
 transition happens at the intermediate values of the coupling where both the weak coupling
 and and strong coupling expansions break down.
 Nevertherless such description is often very useful for phenomenological reasons: for example,
 the walking technicolor models are precisely of this type, since they stay very close to the putative conformal
 fixed point for the long RG time.
 In \KaplanKR\ a tachyon dynamics in the AdS space-time was shown to holographically model this
 type of physics; this has been further studied in \KutasovUQ\ in the context of a particular holographic
 model based on the Tachyon DBI action in AdS.
 The mass of the tachyon is tuned to the critical value (the BF bound) and at the same time the UV
 cutoff is sent to infinity, so that the physical scale measured, for example, by the meson masses, stays fixed.

In this paper we study some phenomenological applications of the model proposed in \KutasovUQ\
which, in turn, was motivated by  the holographic description of the  dynamics of the D3 and D7 branes
intersecting along 2+1 dimensions  \KutasovFR.
We restrict our attention to four space-time dimensions.
In the next Section we investigate the phase diagram of the holographic model at finite temperature and charge density.
We show that the phase transition at finite temperature between the symmetric and the massive phase
is generalized into the phase transition line  in the temperature-charge density plane.
Furthermore, depending on the value of the quartic coefficient in the tachyon potential, the phase transition
line can either stay first order, or possess a critical point where the order of the
phase transition changes from second to first.
This is somewhat similar to the situation with the (conjectural) phase diagram of QCD with massless quarks and  constitutes an interesting prediction for the phase diagram of
strongly interacting fermions.

In Section 3 we explore a possibility of using  the holographic TDBI
model in the context of holographic walking technicolor. We couple
the tachyon bilinear to the gauge fields in the adjoint
representations of $SU(N_f)_L$ and $SU(N_f)_R$ which contain
electroweak gauge group (setting $N_f=2$ and embedding the
electroweak group as $SU(2)\times U(1)\subset SU(2)_L\times SU(2)_R$
constitutes the simplest setup). Tachyon condensate breaks
electroweak  symmetry and generates masses for the W and Z bosons
giving rise to a model  of holographic walking technicolor. We
compute the Peskin-Takeuchi S-parameter for a variety of tachyon
potentials and observe that it is positive and does not go to zero.
In Section 4 we compute the masses of the lightest scalar
technimesons for a certain family of the tachyon potentials and
observe that even though there is no parametrically light
"technidilaton", the lowest lying meson can be an order of magnitude
lighter then the next one.

We conclude in Section 5.
Appendix contains application of the holographic RG to the holographic tachyon TDBI model, where
a picture for the running of the double trace coupling, expected from field theoretic considerations, is
reproduced.

\newsec{Holographic TDBI at finite temperature and chemical
potential}

In this Section we  consider holographic tachyon DBI model at finite temperature and chemical
potential.
We consider AdS-Schwarzshild black hole to account for a
non-vanishing temperature, and we turn on a
background flux of  the $U(1)$ gauge potential
which corresponds to to the finite density in the dual field theory.
We describe the phase with broken conformal symmetry by the
dual picture with non-vanishing tachyon field
in the bulk, while conformally symmetric field theory state
corresponds to the identically vanishing tachyon in the bulk. We compute holographically free energies of both phases and
determine the resulting phase diagram.



Perhaps the future development of the results of this Section will
mostly lie in the realm of condensed matter physics. However, let us
make a slight detour and remind the reader a closely related
problem, a phase diagram of QCD at finite temperature and chemical
potential. (See e.g. \StephanovWX\CasalbuoniRS\ for recent reviews).
The phase structure of QCD is roughly the following.
 If the temperature is low and we
increase the density, then at some value of the density the system
is expected to undergo a first order phase transition to the states
where the hadrons dissociate. At sufficiently large density, the
system gets into the color superconducting phase. In this phase
confined bound state of two quarks goes to Coulomb bound state, in a
process similar to Cooper pairing in the microscopic description of
a superconductor. Increasing the temperature destroys Cooper pairing
mechanism for the quarks, eventually giving rise to a quark-gluon
plasma. This is believed to be a preferred high temperature state
for any values of the chemical potential, however the phase
transition from the hadronic state is first order for larger
densities, but second order for smaller densities (for massless
quarks). As we will see below, we can observe somewhat similar phase
structure for certain TDBI models, though either the orders of first
and second order phase transitions are interchanged or we have two
critical points at which the order of phase transition changes.

Phase transitions in the holographic tachyon DBI at finite
temperature have been studied in \KutasovUQ, which the reader is
encouraged to consult for technical details relevant to the present
Section\foot{ In recent work \AlhoMH\ the phase structure of the
holographic model of QCD in the Veneziano limit has been analyzed at
finite temperature.}. There it has been established that the order
of the phase transition is determined by the behavior of the tachyon
potential for very small values of $T$ (the tachyon field). In the
BKT limit, where the UV cutoff is taken to infinity, with physical
observables held fixed, the solution must have a fixed ratio between
the two asymptotics near the boundary of AdS. The value of the
coefficient in front of the $T^4$ term in the tachyon potential
determines whether increasing the  value of $T$ at the black hole
horizon corresponds to the smaller or larger temperatures. In the
former case, the transition is second order, while in the latter
case it is first order. In the following we repeat this analysis in
the presence of finite density.

Consider finite temperature $AdS_{d+1}$-Schwarzshild metric
\eqn\adsschw{ds^2=r^2\left(-F(r)dt^2+(dx^1)^2+...+(dx^{d-1})^2\right)
+\frac{dr^2}{r^2F(r)}\,,}
where $F(r)=1-\left(\frac{r_h}{r}\right)^d$, and turn on
non-vanishing flux $\dot A_0$. Tachyon-DBI action then takes the
form
\eqn\tdbiaoprdef{S_{TDBI}=-\int _{r_h}^\infty dr\int d^dx
r^{d-1}V(T)\sqrt{1+r^2\dot T^2 F-\dot A_0^2}\,.}
From equation of motion for gauge flux we obtain
\eqn\Azerosol{\dot A_0^2=\frac{\hat d^4(1+ r^2 F\dot T^2)}{
r^{2(d-1)}V^2+\hat d^4}\,,}
where $\hat d(\mu_{ch}, r_h)$ is a constant of integration.
As usual, up to a normalization constant,  ${\hat d}^2$ is proportional to the charge density
of the system.
Due to \Azerosol\ in the leading order in $T$ we obtain
\eqn\muchdhat{\mu _{ch}=\int _{r_h}^\infty \frac{dr\,\hat
d^2}{\sqrt{\hat d ^4+r^{2(d-1)}}}=\frac{\hat d^2}{
(d-2)r_h^{d-2}}\,_2F_1\left(\frac{1}{2}\,,\,\frac{d-2}{2(d-1)}\,,\,\frac{3d-4}{2(d-1)}\,,\,
-\frac{\hat d^4}{r_h^{2(d-1)}}\right)}
Plugging \Azerosol\ in into the action \tdbiaoprdef\
we arrive at
\eqn\tdbiaoonsh{S_{TDBI}=-\int _{r_h}^\infty dr\int
d^dxr^{2(d-1)}V^2\left((1+r^2F\dot T^2)(r^{2(d-1)}V^2+\hat
d^4)^{-1}\right) ^{1/2}\,.}
Introduce dimensionless coordinate, $\rt=r/\hat d^{\frac{2}{d-1}}$,
and dimensionless temperature, $\rt _h=r_h/\hat d^{\frac{2}{d-1}}$.
As a result the action acquires the form
\eqn\tdbidout{S_{TDBI}=-\hat d^{\frac{2d}{d-1}}\int _{\rt _h}^\infty
d\rt\int
d^dx\frac{\rt^{2(d-1)}V^2}{\sqrt{1+\rt^{2(d-1)}V^2}}\sqrt{1+\rt^2FT^{\prime
2}}\,,}
where $F=1-(\rt_h/\rt)^d$ and $T'=\p T/\p \rt$.

Let us define tachyon $T$ value at the horizon, $T_h=T(r_h)$.
Equation of motion for the tachyon field, following from the action
\tdbidout, is
\eqn\rTeqn{\left(\frac{\rt^{2d}FV^2T'}{\sqrt{(1+\rt^2FT^{\prime
2})(1+\rt^{2(d-1)}V^2)}}\right)'-\rt^{2(d-1)}V^2\frac{2+\rt^{2(d-1)}V^2}{(1+\rt^{2(d-1)}V^2)^{3/2}}
\sqrt{1+\rt^2FT^{\prime 2}}\,\p_T\log V=0\,.}
Using \rTeqn\ and imposing the boundary condition $T(\rt=\rt _h)=T_h$, we find
\eqn\Tprrbound{T'(\rt=\rt_h)=\frac{2+\rt_h^{2(d-1)}V^2(T_h)}{d\rt_h(1+\rt_h^{2(d-1)}V^2(T_h))}\p_T\log
V(T_h)\,.}
When $T\sim T_h\ll 1$ and $m\simeq m_{BF}^2=-d^2/4$ we obtain
linearized equation of motion
\eqn\Trlineqn{\left(\frac{\rt^{2d}FT'}{\sqrt{1+\rt^{2(d-1)}}}\right)'+
\frac{d^2\rt^{2(d-1)}}{4}\frac{2+\rt^{2(d-1)}}{(1+\rt^{2(d-1)})^{3/2}}T=0}
and boundary conditions
\eqn\linbcondyone{T(\rt=\rt_h)=T_h\,,\quad\quad T'(\rt=\rt_h)
=-\frac{dT_h}{4\rt_h}\frac{2+\rt_h^{2(d-1)}}{1+\rt_h^{2(d-1)}}\,.}
Near the boundary $\rt\rightarrow\infty$ behavior of $T(\rt)$ is
given by equation
\eqn\nbteqt{T''+\frac{d+1}{\rt}T'+\frac{d^2}{4\rt^2}T=0\,.}

Let us now specialize to $d=4$ case. Near-boundary behavior is then
described by equation
\eqn\nbteqtf{T''+\frac{5}{\rt}T'+\frac{4}{\rt^2}T=0\,.}
which is solved by
\eqn\nbTtr{T(\rt)\simeq \frac{1}{\rt^{2}}\left(c_1\log\rt
+c_2\right)\quad\Rightarrow\quad
T(r)=\frac{1}{r^{2}}\left(c_1\log\frac{r}{\hat
d^{2/3}}+c_2\right)\,.}
Let us denote
\eqn\gdefinition{   g={\hat d}^{2/3}   }
 The constants $c_1$ and $c_2$ can be
determined by solving equation of motion \rTeqn\ numerically. If we
consider instead linearized equation \Trlineqn\ in BKT limit with
boundary conditions \linbcondyone, we obtain $c_2/c_1$, which is a
function of $\rt_h=r_h/g$. In the case of vanishing temperature and
vanishing chemical potential the near-boundary behavior of tachyon
field is given by
\eqn\Nbdefmu{T(r)=\frac{1}{r^{2}}\left(C_1\log\frac{r}{\mu}
+C_2\right)}
Clearly it must be the same as \nbTtr\ . Matching these equations,
we obtain
\eqn\Matchcoct{{\frac{g}{\mu}=C_0
\exp\left(\Xi\left(\frac{r_h/\mu}{g/\mu}\right)\right)}\,,}
where we have denoted $C_0=e^{-C_2/C_1}$ and $\Xi=c_2/c_1$. Equation
\Matchcoct\ can be solved numerically, which gives critical values
of temperature $r_h$ and $g$ measured in units of $\mu$.
The result appears in figure 1. We have checked that when ${\hat
d}=0$ the critical temperature is equal to $2\,C_0$, which is a
correct limiting value \KutasovUQ.

\ifig\loc{Phase diagram for conformal phase transition in $(g/\mu,\;
r_h/\mu)$ plane. Order of phase transition changes at the point
$\tilde r_h\equiv r_h/g=0.75$. Blue part of the curve describes
second order phase transition and red part of the curve describes
first order phase transition.} {\epsfxsize3.3in\epsfbox{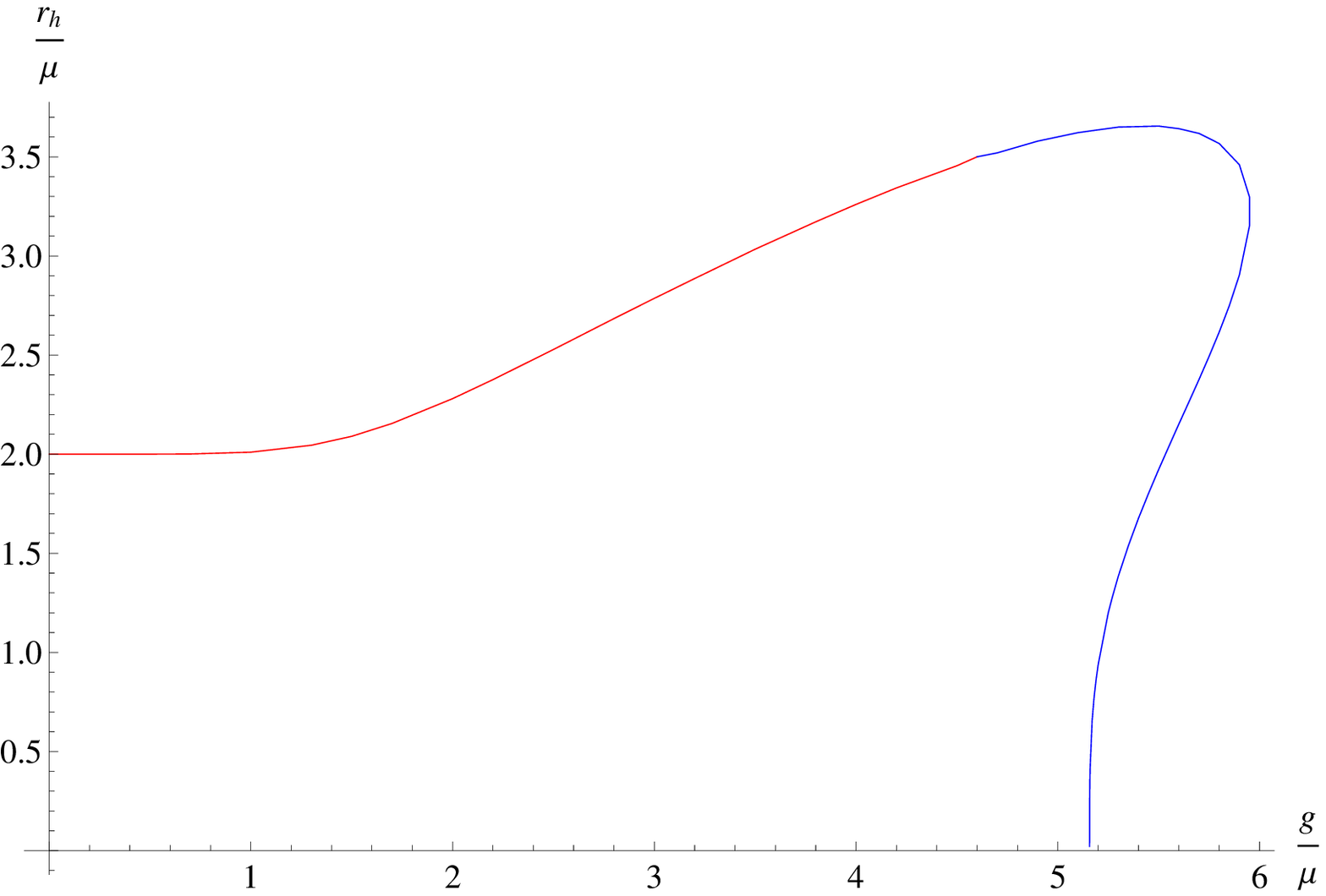}}
\bigskip

To determine which state in the canonical ensemble is preferred, we
need to compare the free energies. Similarly to \KutasovUQ, we focus
on the near-critical region, where tachyon field is either vanishing
or small. The difference in free energies between non-vanishing
tachyon and vanishing tachyon phases is given by
\eqn\FFdefone{\FF (r_h,\hat d)=S_{TDBI}(T\equiv 0)-S_{TDBI}(T)\,,}
where the last term in the r.h.s. is evaluated on the solution,
satisfying $T(r=r_h)=T_h$ boundary condition. Due to $V(0)=1$ one
obtains, using \tdbidout\
\eqn\FFdef{\FF(r_h,\hat d)=\hat d^{8/3}\int  _{\rt_h}^\infty
d\rt\int d^4x\rt^6\left(\frac{V^2}{\sqrt{1+\rt^6V^2}}\sqrt{1+
\rt^2FT^{\prime 2}}-\frac{1}{\sqrt{1+\rt^6}}\right)\,.}
We will need the form  of the tachyon potential near $T=0$:
\eqn\Vexp{V(T)=1+\frac{1}{2}m^2T^2+\frac{a}{4}T^4+\cdots\,,}
where $m^2\simeq m_{BF}^2=-4$ and $a$ is the coefficient of the quartic
term which, as explained in \KutasovUQ, determines the order of the phase
transition in the case of vanishing density.
Below we will see that at finite density the situation is more subtle, and the
first order phase transition line can join the second order phase transition
line at a critical point, provided the value of $a$ is chosen accordingly.

In the BKT limit we have $T\sim T_h\ll 1$ and $m^2=m_{BF}^2=-4$. We
compute \FFdef\ up to the fourth order in $T_h$, \eqn\FFdectwf{\FF(
r_h,\hat d)=\FF_2( r_h,\hat d)+\FF_4( r_h,\hat d)+\cdots\,,} where

\eqn\gzeroca{\eqalign{  \FF_2 & =\hat d^{8/3}\int _{\rt_h}^\infty
d\rt\int d^4x\frac{1}{2\sqrt{1+\rt^6}}\left(\rt^8FT^{\prime
2}-4\rt^6\frac{2+\rt^6}{1+\rt^6}T^2\right)\cr \FF_4 & =-\hat
d^{8/3}\int _{\rt_h}^\infty d\rt\int
d^4x\frac{\rt^6}{8(1+\rt^6)^{5/2}}\left(F^2T^{\prime
4}\rt^4(1+\rt^6)^2 +\right.\cr &+\left.8F\rt^2T^2T^{\prime
2}(1+\rt^6)(2+\rt^6)+
              2T^4(8(\rt^6{-}2){-}a(1+\rt^6)(2+\rt^6))\right)}}

The quadratic terms vanish on shell, up to the boundary term,
which also vanishes, because
\eqn\FTbc{F(\rt=\rt_h)=0\,,\quad\quad T(\rt=\infty)=0\,.}
We solve numerically equation \Trlineqn\ with boundary conditions
\linbcondyone, for each particular $\rt_h=r_h/g$. This gives us
$T=T_1$, which is a solution of the first order in $T_h$. The first
correction to this solution is obtained when we take into account
quartic in $T_h$ terms in the action for $T$, and therefore the
corrected solution is $T=T_1+T_3$, where $T_3$ is of the third order
in $T_h$. Therefore we need to compute in the leading order
\eqn\FFtwofour{\FF(T_1+T_3)=\FF_2(T_1+T_3)+\FF_4(T_1+T_3)\,.}
For brevity let us rewrite \gzeroca\ as
\eqn\gzerocash{\eqalign{  \FF_2 & =\int dr[\alpha(r)T^2+\beta
(r)T^{\prime 2}]\cr \FF_4 & =\int dr[a(r)T^4+b(r)T^2T^{\prime
2}+c(r)T^{\prime 4}]}}
Let us use integration by parts to bring $\FF_{2,4}$ to the form
\eqn\gzerocash{\eqalign{  \FF_2 & =\int dr\, T[\alpha T-(\beta
T^\prime )']\equiv \int dr\, TP_1\cr \FF_4 & =\int dr\,
T\left[aT^3+\frac{b}{2}TT^{\prime
2}-\left(\frac{b}{2}T'T^2\right)'-(cT^{\prime 3})'\right]\equiv\int
dr\, TP_3 }}
where $P_{1,3}$ are polynomials of the $T,\,T',\,T''$ of the degree
specified by the subscript. From the variation
\eqn\gzerocvar{  \delta\FF =2\int dr\, \delta T[\alpha T-(\beta
T^\prime )']+4\int dr\, \delta T\left[aT^3+\frac{b}{2}TT^{\prime
2}-\left(\frac{b}{2}T'T^2\right)'-(cT^{\prime 3})'\right]}
we obtain equation of motion
\eqn\Tpoleq{2P_1(T_1+T_3)+4P_3(T_1+T_3)=0\,,}
which we can solve perturbatively as
\eqn\Tpoleqs{P_1(T_1)=0\,,\quad P_1(T_3)+2P_3(T_1)=0\,.}
Using these equations in the expansion of \FFtwofour
\eqn\FFtwofoure{\FF=\int
dr(T_1{+}T_3)P_1(T_1{+}T_3){+}(T_1{+}T_3)P_3(T_1{+}T_3){=}\int dr
T_1P_1(T_3){+}T_3P_1(T_1){+}T_1P_3(T_1)+\cdots}
we obtain
\eqn\FFpert{\FF(T)\simeq -\int drT_1P_3(T_1)=-\FF_4(T_1)\,.}

We then evaluate quartic terms, $\FF_4(\hat d,a)$, on the
numerically found solution $T_1$. Equation $\FF_4(\rt_h,\,a)=0$
gives values of ratio $r_h/g=\rt_h$ for each particular $a$ at which
order of phase transition changes. This equation is valid only for
those values of $r_h$ and $\hat d$ which are close to critical ones.
We solve this equation numerically for each particular value of the
parameter $a$, that is we find $\rt_h ^{(c)}(a)$. The result is
plotted in figure 2. Notice that when $\hat d$ is sent to zero,
$\rt_h$ goes to infinity, and the special value $a\simeq 6.47$
becomes the same as in the case of vanishing chemical potential
\KutasovUQ. Also notice that when $6.47\leq a\leq 7.03$ there are
two points $\rt_h$ at which the order of phase transition changes.

\ifig\loc{The ratio $\rt_h=r_h/g$ at which the order of phase
transition changes, as a function of UV parameter $a$. It is
determined by the sign of $\FF_4$ in the conformal symmetry broken
phase. On the left side of the curve $\FF_4 <0$ and phase transition
is of the first order, on the right side $\FF_4
>0$ and phase transition is of the second order.}
{\epsfxsize3.3in\epsfbox{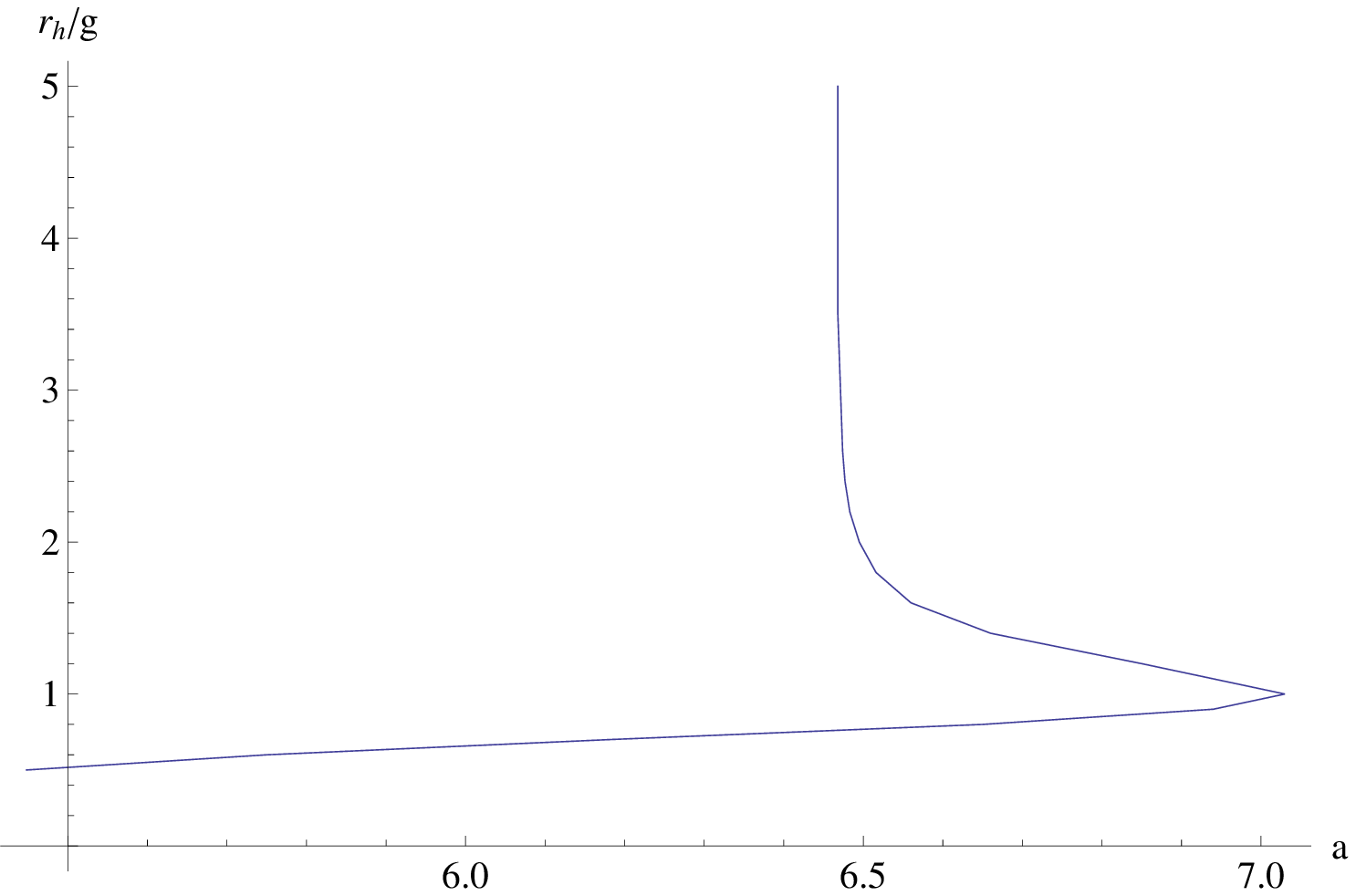}}

In figure 1 we have taken $a=6.41$ for which phase transition is the
second order one for $\frac{r_h}{g}<0.75$ and the first order one
for $\frac{r_h}{g}>0.75$. This corresponds respectively to the blue
and red parts of the phase transition curve in figure 1. \foot{One
may use the top-down approach based on the $D3-D7$ system to derive
the phase diagram of $N=4$ super Yang-Mills coupled to $N=2$ matter
at finite temperature and chemical potential. It also exhibits the
phase transition of the second order at small temperatures. See
\EvansAT\ for a recent discussion.}

The other option is to take the value of $a$ at which we have two
critical points where the order of phase transition changes. Then
for the temperature bellow some critical value, $\tilde r_h^{(c)}<
r_h^{(c,1)}$ we have second order phase transition, for $\tilde
r_h^{(c,1)}<\rt_h^{(c)}<\rt_h^{(c,2)}$ we have first order order
phase transition, and finally for $\rt_h^{(c)}>\rt_h^{(c,2)}$ we
have second order phase transition. The critical point
$\rt_h^{(c,2)}$ therefore resembles the one in the QCD phase
diagram.

As emphasized in \KutasovUQ, the behavior of the free energy for small
values of the tachyon condensate determines the order of the phase transition, provided
the phase diagram has a simple form. This was the case in all examples
studied in \KutasovUQ. We believe this remains true once the finite density
is turned on, but to show this some further  numerical work is necessary.

\newsec{S parameter}

\subsec{ Review of technicolor and $S,T,U$ parameters}

Consider the system of $2$ techniquark matter fields $(\tilde
u,\tilde d)$ with color charges, transforming in fundamental
representation of the gauge group $SU(N_c)$. Quark fields are
coupled to gauge field in the adjoint representation of the gauge
group. In the ultraviolet regime these quarks are massless, and
therefore the system possesses the $SU(2)_L\times SU(2)_R$ chiral
symmetry. Therefore we can couple the doublet of left quarks
$Q=(\tilde u_L\,,\; \tilde d_L)$ to bosons of weak gauge group
$SU(2)_L$, leaving $2$ right quarks $\tilde u_R$ and $\tilde d_R$ in
the singlet representation sector of the weak gauge transformations.
We also give each quark field the hypercharge $Y$, characterizing
its representation under the action of the $U(1)_Y$ gauge group.


We look at introduced $2$ quarks as a set of strongly-interacting
fermionic fields of the physics beyond the Standard Model. At some
energy scale due to the strong interaction these quarks may form a
chiral condensate, breaking the chiral symmetry down to
$SU(2)_{diag}$. \foot{It was shown in \ColemanMX\ that under general
assumptions in large-$N_c$ chromodynamics the chiral symmetry breaks
spontaneously.} In the vacuum with spontaneously broken chiral
symmetry the $2$ quarks acquire a  mass.  In the technicolor models
the phenomenon of the chiral symmetry breaking via techniquark
condensation is used to explain the spontaneous electro-weak
symmetry breaking, realized therefore as a dynamical symmetry
breaking. Furthermore, the extended technicolor models combine these
$2$ techniquarks with SM matter fields in some specific multiplets
in such a way that condensation of techniquarks gives masses to SM
matter fields. In the simplest technicolor models such quartic
fermionic terms,  generated at high scale $\Lambda_{ETC}$, lead to
flavor changing neutral currents matrix elements that are way above
experimental bounds. A walking technicolor models, where the system
spends a long RG time in a vicinity of a putative RG fixed point and
the anomalous dimension of the technimeson condensate $\gamma\simeq
1$, has been proposed to alleviate this problem. (See \PiaiMA\ for a
recent review of walking technicolor and references therein). The
theory is necessarily strongly coupled, and it is natural to use
holography in this context.

To create a possibility for experimental tests of theories
describing physics beyond the Standard Model, Pesking and Takeuchi
\refs{\PeskinZT,\PeskinSW} introduced dimensionless parameters
$S\,,\;T\,,\;U$, measuring an impact of a hidden sector of heavy
beyond-SM fundamental matter fields coupled to electro-weak gauge
bosons. They argued (following \refs{\KennedySN}) that the most
important impact arises from oblique corrections: vacuum
polarization diagrams, which renormalize gauge boson propagators.
Peskin-Takeuchi parameters are expressed via these vacuum
polarization amplitudes, and we will review their argument in a
greater detail bellow. For each beyond-SM theory we therefore may
compute $S\,,\;T\,,\;U$ parameters and see whether the results lie
within the boundaries set by the deviation of experimental data from
Standard Model predictions.


The quantum corrections of matter fields to the propagators of the
SM gauge fields come from the vacuum polarization amplitudes
\eqn\vpolamp{\int d^4x e^{iq\cdot x}\langle J^{\mu}_ a(x)J^{\nu}_
b(0)\rangle =-i\left(g^{\mu\nu}-\frac{q^\mu q^\nu}{q^2}\right)\Pi
_{ab}(q^2)\,,}
where $a,b=1,2,3,Q$ and we are assuming mostly plus signature of the
metric. The expression \vpolamp\ should be computed for the matter
fields of the SM and for the hidden matter sector of the beyond-SM
physics.
For weak currents $J_i\,,i=1,2$, weak isospin
current $J_3$ and electromagnetic current $J_Q$ we have the vacuum
polarization amplitudes
\eqn\vpolampp{\Pi_{11}\,,\;\Pi_{22}\,,\;\Pi_{33}\,,\;\Pi_{3Q}\,,\;\Pi
_{QQ}\,.}
If we know these amplitudes, then using expression for the
electroweak interaction Lagrangian
\eqn\elwintl{L=\frac{e}{\sqrt{2}s}\left(W_\mu ^+J_+^\mu +W_\mu
^-J_-^\mu\right) +\frac{e}{sc}Z_\mu \left(J^\mu _3
-s^2J_Q^\mu\right)+eA_\mu J^\mu _Q}
we can obtain 1PI self-energies for the electroweak gauge bosons,
and 1PI mixing for $Z$-boson and photon,
\eqn\pol{\Pi_{AA}=e^2\Pi_{QQ}\,,\quad\Pi_{ZA}=\frac{e^2}{sc}(\Pi_{3Q}-s^2\Pi_{QQ})\,,...}
Then with the help of Schwinger-Dyson equations we can derive full
quantum propagators for the electroweak gauge fields.

Now, in
the interaction Lagrangian \elwintl\ we have parameters $e$ and
\eqn\defs{ s^2\equiv\sin ^2\theta_W=1-\frac{m_W^2}{m_Z^2}.}
 Quantum
corrections due to the vacuum polarization amplitudes boil down to
the renormalization of these parameters,
\eqn\estar{e_\star^2(q^2)\equiv\frac{e^2}{1-e^2\Pi_{QQ}(q^2)}\,,}
\eqn\sstar{s_\star^2(q^2)\equiv
s^2-sc\frac{\Pi_{ZA}(q^2)}{q^2-\Pi_{AA}(q^2)}\,.}
%

Then, the renormalized parameter $s_\star$ enters the measured
left-right $Z$-decay asymmetry,
\eqn\ALRss{A_{LR}(q^2)=\frac{2(1-4s_\star^2)}{1+(1-4s_\star^2)^2}\,,}
and therefore the renormalization of the gauge fields propagators
(coming mainly as the oblique corrections due to loops of heavy
fermions) can be measured experimentally.

Let us also define $\theta _0$ as
\eqn\thetand{\sin (2\theta _0)=\sqrt{\frac{4\pi \alpha_{\star,
0}(m_Z^2)}{\sqrt{2}G_Fm_Z^2}}\,.}
Here $m_Z$ and $G_F$ are experimentally measured. And
$\alpha_{\star,0}(m_Z^2)$ is a running electromagnetic coupling,
which is computed due to known physics up to $q^2=m_Z^2$ scale. The
running starts from the measured $\alpha (q^2=0)=e^2/(4\pi)$.


The renormalization comes from SM and from physics beyond the SM. In
the SM the most important contribution comes from $t$-quark loops
(see \eg\ \PeskinEV\ Chapter 21),
\eqn\smcorone{s^2-s_\star^2=-\frac{3\alpha c^2}{16\pi
s^2}\frac{m_t^2}{m_Z^2}\,,}
\eqn\smcortwo{s_\star ^2-s_0^2=-\frac{3\alpha}{16\pi
(c^2-s^2)}\frac{m_t^2}{m_Z^2}\,.}

Let us now describe quantum corrections due to vacuum polarization
diagrams of beyond-SM physics. First of all for heavy fermion we
can expand vacuum polarization amplitudes around $q^2=0$,
\eqn\vpolexpo{\Pi _{QQ}(q^2)=q^2\Pi_{QQ}'(0)\,,\quad
\Pi_{3Q}(q^2)=q^2\Pi_{3Q}'(q^2)\,,}
\eqn\vpolexpt{\Pi_{33}(q^2)=\Pi_{33}(0)+q^2\Pi_{33}'(0)\,,}
\eqn\vpolexptr{\Pi_{11}(q^2)=\Pi_{11}(0)+q^2\Pi_{11}'(0)\,,}
where prime denotes differentiation w.r.t. $q^2$ and we have made
use of the fact that Ward identity for electromagnetic field ensures
$\Pi_{QQ}(0)=0$ and $\Pi_{3Q}(0)=0$. Also we have
$\Pi_{11}=\Pi_{22}$. We have therefore six parameters defining
vacuum polarization amplitudes of heavy fermions. We make a
renormalization, fixing values of three well-measured parameters,
which are $\alpha$, $G_F$ and $m_Z$. Three parameters which are left
are free of UV divergencies, and we combine these into
\eqn\PTpardefo{\alpha
S=4e^2\left(\Pi_{33}'(0)-\Pi_{3Q}'(0)\right)\,,}
\eqn\ptpardeft{\alpha
T=\frac{e^2}{s^2c^2m_Z^2}\left(\Pi_{11}(0)-\Pi_{33}(0)\right)\,,}
\eqn\ptpardeftr{\alpha
U=4e^2\left(\Pi_{11}'(0)-\Pi_{33}'(0)\right)\,.}

In addition to SM corrections \smcorone\ and \smcortwo\ we can write
down contribution of beyond-SM physics via these parameters:
\eqn\npcontro{\frac{m_W^2}{m_Z^2}-c_0^2=\frac{\alpha
c^2}{c^2-s^2}\left(-\frac{1}{2}S+c^2T+\frac{c^2-s^2}{4s^2}U\right)\,,}
\eqn\npcontrtw{s_\star
^2-s_0^2=\frac{\alpha}{c^2-s^2}\left(\frac{1}{4}S-s^2c^2T\right)\,.}
Thus we explicitly constructed a set of experimentally measured
quantities, quantum corrections to which may be separately computed
from the SM, \smcorone, \smcortwo, and from a hidden sector,
\npcontro, \npcontrtw, with the latter being expressed via
Peskin-Takeuchi parameters.

Let us use vector and
axial-vector isospin currents
\eqn\vau{J^\mu _V=\bar\psi\gamma ^\mu\tau _3\psi\,,\quad\quad J^\mu
_A=\bar\psi\gamma ^\mu\gamma^5\tau _3\psi\,,}
to express left isospin current as
\eqn\vautwo{J^\mu _3=\frac{1}{2}(J^\mu _V-J^\mu _A)\,.}
Consider also electromagnetic current, expressed via isospin and
hypercharge currents in a usual way,
\eqn\elmcur{J^\mu _Q=J^\mu _V+\frac{1}{2}J_Y^\mu\,.}
Assuming the conservation of parity by technicolor interactions we
can express isospin current correlator via vector and axial vector
isospin correlators, $\Pi_{33}=\frac{1}{4}(\Pi_{VV}+\Pi_{AA})$. We
also note that due to isospin conservation $\langle J_3 J_Y\rangle
=0$ (otherwise in technicolor models there would have been a
preferred isospin direction), we obtain
$\Pi_{3Q}=\frac{1}{2}\Pi_{VV}$. Therefore
\eqn\sparvaexp{S=-4\pi (\Pi_{VV}'(q^2)-\Pi _{AA}'(q^2))|_{q^2=0}\,.}

The holographic
tachyon DBI was introduced in \KutasovUQ\ to describe a system of
strongly interacting fermions, which can be made into the walking technicolor theory.
However for the purpose of computing the S-parameter and technimeson masses,
we do not even need to specify that the holographic TDBI model describes strongly interacting fermions.
Instead,  it is sufficient to treat the holographic model as a black box, which
produces two-point functions for the vector and axial currents and
features spontaneous breaking of the axial symmetry. Then, these
currents are coupled to the SM gauge fields to produce spontaneous
symmetry breaking of the electroweak gauge group. The resulting
contribution to the S-parameter is given by \sparvaexp\ and is
computed below.

\subsec{Computation of the $S$ parameter from the tachyon DBI action}

As pointed out above, the holographic tachyon DBI theory provides
a natural model of the walking technicolor scenario.
The important feature of the holographic approach is
that we can  isolate the impact of beyond-SM sector of the
theory. For this purpose we just have to consider a corresponding
set of fields in the bulk, and study its classical dynamics\foot{Previous work dedicated to holographic technicolor and $S$
parameter includes
\refs{\HongSI\HirnNT\PiaiHY\CaroneWJ\AgasheMC\CaroneMD\CasalbuoniXN\HirayamaHZ\CaroneRX\FabbrichesiGA\SanninoHA\HabaNZ\DietrichNI\DietrichUP\NunezWI
\MintakevichWZ\FukanoZM\KitazawaKR\DietrichAF\BelitskyFJ\CaroneCP\ReeceXJ\HabaHU\AnguelovaQH
\RoundNH\AnguelovaBC\SondergaardPS\LevkovYK\AnguelovaKA
\CaroneCD\LawranceCG-\ElanderFK}; see also
\refs{\JarvinenQE\AlvaresKR\MatsuzakiMK\EvansJX\MatsuzakiXX\ChackoVM-\BellazziniVZ} for recent related
work. }.

We need to construct a dual to a strongly interacting theory with the $SU(2)_L\times SU(2)_R$ global symmetry in the UV.
The global
currents $j_\mu^{(L)}$ and $j_\mu ^{(R)}$ give rise to the bulk fields
$A_M ^{(L,R)}$, living in adjoint of $SU(2)_{L,R}$, with gauge
transformations
\eqn\Altr{A_M^{(L)}\rightarrow U_LA_M^{(L)}U_L^\dagger
+i\p_MU_LU_L^\dagger\,,\quad\quad A_M^{(R)}\rightarrow
U_RA_M^{(R)}U_R^\dagger +i\p_MU_RU_R^\dagger\,.}
Tachyon field $T(r,x)$ lives in bi-fundamental of $SU(2)_L\times
SU(2)_R$, \ie\ its gauge transformations are given by
\eqn\Ttr{T\rightarrow U_LTU_R^\dagger\,.}
Tachyon action with $SU(2)_L\times SU(2)_R$ local symmetry in the
bulk is then
\eqn\tachlrbd{S=-\int d^4xdr\,\Tr\,
V(|T|)\left(\sqrt{-G^{(L)}}+\sqrt{-G^{(R)}}\right)\,,}
where
\eqn\GLGR{G^{(R)}_{MN}=G_{MN}+F_{MN}^{(R)}\,,\quad
G_{MN}=g_{MN}+(D_{(M}T)^\dagger D_{N)}T\,,\quad G^{(R)}=\det\,
G_{MN}^{(R)}\,,}
and similar for the left; and covariant derivative of tachyon
field is given by
\eqn\DLRT{D_MT=\p_M T+iA_M^{(L)}T-iTA_M^{(R)}\,.}
Similar actions for the tachyon have been introduced in \CaseroAE.

We have $A_M^{(L,R)}=(A_r^{(L,R)},A_\mu ^{(L,R)})$ and we partly fix
the gauge symmetry putting
\eqn\gsymbf{A_r^{(L)}=0\,,\quad\quad A_r^{(R)}=0\,.}

Let us introduce gauge fields in the bulk, dual to vector and axial
currents on the boundary:
\eqn\varedef{A^{(L)}_M=\frac{1}{2}(V_M-A_M)\,,\quad\quad
A^{(R)}_M=\frac{1}{2}(V_M+A_M)}

Suppose we have background tachyon field $T(r)=\langle T(r)\rangle
I$, with real-valued vacuum average $\langle T(r)\rangle =T_0(r)$,
satisfying equation of motion at vanishing gauge fields,
\eqn\tneq{\frac{d}{dr}\left(\frac{r^5\dot T_0}{\sqrt{1+r^2\dot
T_0^2}}\right)=\frac{r^3\p_T\log\,V(T_0)}{\sqrt{1+r^2\dot T_0^2}}}
%
%
%

Such a background tachyon field breaks the symmetry down to
$SU(2)_{diag}$, which means $U_L=U_R$. Its non-zero covariant
derivative components, due to the gauge choice \gsymbf\ and
definition \varedef\ are (the fact that $T$ couples only to axial
field $A$ means that axial symmetry is broken)
\eqn\dtnvc{D_r T=\dot T_0\,I\,,\quad\quad D_\mu T=-iA_\mu T_0\,.}
In what follows we consider the case of just one flavor of quark
fields. The results can be generalized to arbitrary number of
flavors, because for the holographic computation of two-point
functions higher order non-abelian terms in gauge field Lagrangian
do not play any role. We therefore have
\eqn\gmns{G_{MN}=g_{MN}+\p_{M}T_0\p_{N}T_0+A_MA_NT_0^2\,.}
Let us denote for brevity
\eqn\gnmns{\GG_{MN}=g_{MN}+\p_{M}T_0\p_{N}T_0={\rm
diag}\,\left(-r^2,\;r^2,\;r^2,\;r^2,\;\frac{1+r^2\dot
T_0^2}{r^2}\right)\,,}
and let us write down an inverse matrix to \gnmns\
\eqn\gnmninvs{{\cal M}^{MN}\equiv (\GG ^{-1})^{MN}={\rm
diag}\,\left(-\frac{1}{r^2},\;\frac{1}{r^2},\;\frac{1}{r^2},\;\frac{1}{r^2},\;\frac{r^2}{1+r^2\dot
T_0^2}\right)\,.}
We also denote
\eqn\gdetdef{\sqrt{-G}=\sqrt{-\det\,||\,G_{MN}\,||}\,,\quad\quad
\sqrt{-G_0}=\sqrt{-\det\,||\,\GG_{MN}\,||}=r^3\sqrt{1+r^2\dot
T_0^2}\,,}
\eqn\kmndef{K_{MN}=G_{MN}-\GG_{MN}=A_MA_NT_0^2\,.}

Up to second order in $A$ we expand
\eqn\ddetG{\sqrt{-G}=\sqrt{-G_0}\,\exp\left(\frac{1}{2}\tr\log(1+{\cal
M}K)\right)=r^3\sqrt{1+r^2\dot
T_0^2}\left(1+\frac{T_0^2}{2r^2}\eta^{\mu\nu}A_\mu A_\nu\right)\,.}

Expanding action \tachlrbd\ up to second power of gauge fields and
replacing left and right gauge fields with vectors and axials we get

\eqn\gzeroca{\eqalign{  S  & =-\int
d^4xdrV(T_0)\sqrt{-G}\left(2+\frac{1}{4}(G^{-1})^{M_1M_2}(G^{-1})^{N_1N_2}\left(
F_{M_1N_1}^{(L)}F_{M_2N_2}^{(L)}+F_{M_1N_1}^{(R)}F_{M_2N_2}^{(R)}
\right)\right)\cr
              &{=}{-}\int d^4xdrV(T_0)r^3\sqrt{1{+}r^2\dot
T_0^2}[2{+}\frac{T_0^2}{r^2}\eta^{\mu\nu}A_\mu A_\nu \cr
&\qquad\qquad+
\frac{1}{8}\CM^{M_1M_2}\CM^{N_1N_2}(
F_{M_1N_1}^{(V)}F_{M_2N_2}^{(V)}{+}F_{M_1N_1}^{(A)}F_{M_2N_2}^{(A)}
)]}}

Using expression \gnmninvs\ for $\CM$ and throwing away what is
independent of gauge fields we proceed to
\eqn\tgmn{\eqalign{S&=-\int d^4xdrV(T_0)r^3\sqrt{1+r^2\dot
T_0^2}[\frac{1}{4(1+r^2\dot T_0^2)}\eta^{\mu\nu}(\dot V_\mu \dot
V_\nu +\dot A_\mu \dot A_\nu)\cr &+\frac{1}{8r^4}\eta^{\mu\nu}\eta
^{\lambda\rho}(F_{\mu\lambda}^{(V)}F_{\nu\rho}^{(V)}+F_{\mu\lambda}^
{(A)}F_{\nu\rho}^{(A)})+\frac{T_0^2}{r^2}\eta^{\mu\nu}A_\mu
A_\nu]\,.}}
We now go to momentum representation,
\eqn\ft{V_\mu
(x,r)=\int\frac{d^4q}{(2\pi)^{2}}V_{\mu}(q,r)e^{-iq_\lambda
x^\lambda}\,,\quad\quad A_\mu
(x,r)=\int\frac{d^4q}{(2\pi)^{2}}A_{\mu}(q,r)e^{-iq_\lambda
x^\lambda}\,,}
which results in
\eqn\tgmnmr{\eqalign{S&=-\int d^4qdrV(T_0)r^3\sqrt{1+r^2\dot
T_0^2}[\frac{1}{4(1+r^2\dot T_0^2)}\eta^{\mu\nu}(\dot V_\mu \dot
V_\nu +\dot A_\mu \dot A_\nu)\cr
&+\frac{q^2}{4r^4}\left(V_{\mu}V_\nu\left(\eta^{\mu\nu}-\frac{q^\mu
q^\nu}{q^2}\right)+A_{\mu}A_\nu
\left(\eta^{\mu\nu}\left(1+\frac{4T_0^2r^2}{q^2}\right)-\frac{q^\mu
q^\nu}{q^2}\right)\right)]\,,}}
where all squared gauge fields are just a short notation for
$q$-mode and $-q$-mode product.

Let us split radial and momentum dependence as follows:
\eqn\rmsplit{V_\mu (q,r)=v_\mu (q)v(q,r)\,,\quad\quad A_\mu
(q,r)=a_\mu (q)a(q,r)\,.}
(We can use residual gauge symmetry to gauge-fix $q^\mu V_\mu
(q,\Lambda)=q^\mu A_\mu (q,\Lambda)=0$.) Let us also split the
action \tgmnmr\ to axial and vector parts:
\eqn\tgmngauge{S=S_V+S_A\,,}
where
\eqn\tgmngv{\eqalign{S_V&=-\frac{1}{4}\int
d^4qdr\frac{r^3}{\sqrt{1+r^2\dot T_0^2}}V(T_0)v_\mu (q)v_\nu
(-q)\times \cr &\times\left(\dot v_q(r)\dot
v_{-q}(r)\eta^{\mu\nu}+\frac{q^2(1+r^2\dot
T_0^2)}{r^4}\left(\eta^{\mu\nu}-\frac{q^\mu
q^\nu}{q^2}\right)v_q(r)v_{-q}(r)\right)\,,}}
\eqn\tgmnga{\eqalign{S_A&=-\frac{1}{4}\int
d^4qdr\frac{r^3}{\sqrt{1+r^2\dot T_0^2}}V(T_0)a_\mu (q)a_\nu
(-q)\times \cr &\times\left(\dot a_q(r)\dot
a_{-q}(r)\eta^{\mu\nu}+\frac{q^2(1+r^2\dot
T_0^2)}{r^4}\left(\eta^{\mu\nu}\left(1+\frac{4T_0^2r^2}{q^2}\right)-\frac{q^\mu
q^\nu}{q^2}\right)a_q(r)a_{-q}(r)\right)\,.}}
We are interested in transverse components of gauge fields:
\eqn\trproj{v_\mu^T (q)=P_{\mu\lambda}\eta^{\lambda\nu}v_\nu
(q)\,,\quad a_\mu ^T(q)=P_{\mu\lambda}\eta^{\lambda\nu}a_\nu
(q)\,,\quad\quad P_{\mu\nu}=\eta_{\mu\nu}-\frac{q_\mu
q_\nu}{q^2}\,,}
which are described by
\eqn\tgmngv{\eqalign{S_V^T&=-\frac{1}{4}\int
d^4qdr\frac{r^3}{\sqrt{1+r^2\dot T_0^2}}V(T_0)v_\mu ^T(q)v_\nu
^T(-q)\eta^{\mu\nu}\times\cr &\times\left(\dot v_q(r)\dot
v_{-q}(r)+\frac{q^2(1+r^2\dot
T_0^2)}{r^4}v_q(r)v_{-q}(r)\right)\,,}}
\eqn\tgmnga{\eqalign{S_A^T&=-\frac{1}{4}\int
d^4qdr\frac{r^3}{\sqrt{1+r^2\dot T_0^2}}V(T_0)a_\mu ^T(q)a_\nu
^T(-q)\eta^{\mu\nu}\times\cr &\times\left(\dot a_q(r)\dot
a_{-q}(r)+\frac{q^2(1+r^2\dot
T_0^2)}{r^4}\left(1+\frac{4T_0^2r^2}{q^2}\right)a_q(r)a_{-q}(r)\right)\,,}}
Corresponding equations of motion are
\eqn\vtreq{\ddot v_q(r)+\frac{\sqrt{1+r^2\dot
T_0^2}}{r^3V(T_0)}\frac{d}{dr}\left(\frac{r^3V(T_0)}{\sqrt{1+r^2\dot
T_0^2}}\right)\dot v_q(r)-\frac{q^2(1+r^2\dot
T_0^2)}{r^4}v_q(r)=0\,,}
\eqn\atreq{\ddot a_q(r)+\frac{\sqrt{1+r^2\dot
T_0^2}}{r^3V(T_0)}\frac{d}{dr}\left(\frac{r^3V(T_0)}{\sqrt{1+r^2\dot
T_0^2}}\right)\dot a_q(r)-\frac{q^2(1+r^2\dot
T_0^2)}{r^4}\left(1+\frac{4T_0^2r^2}{q^2}\right)a_q(r)=0\,.}
We see that if there is no tachyon background, then equations of
motion for vector and axial vector fields become the same.

We must ensure that near-horizon behavior of vector and axial vector
fields is regular. The precise boundary conditions in the bulk depend strongly on the tachyon background.
Below
we consider concrete tachyon potentials and determine the corresponding boundary conditions.
We also require
\eqn\UVsource{v(q,r=\infty)=1\,,\quad\quad a(q,r=\infty)=1\,.}
We solve equations of motion for $v(q,r)$ and $a(q,r)$ with these
boundary conditions and plug the solutions into \tgmngv\ and
\tgmnga\ . As a result we obtain (recall that at the boundary
tachyon field vanishes)
\eqn\tgmngvos{S_V^{on-shell}=-\frac{1}{4}\int
d^4q\Lambda^3\eta^{\mu\nu} v_\mu ^T(q)v_\nu ^T(-q)\dot
v(q,\Lambda)\,,}
\eqn\tgmngaos{S_A^{on-shell}=-\frac{1}{4}\int
d^4q\Lambda^3\eta^{\mu\nu}a_\mu ^T(q)a_\nu ^T(-q)\dot
a(q,\Lambda)\,.}

Due to AdS/CFT correspondence
\eqn\coradscft{i\int d^4xe^{iqx}\langle j^\mu _V(x)j^\nu
_V(0)\rangle=\frac{\delta ^2S_V^{on-shell}}{\delta v_\mu ^T(q)\delta
v_\nu^T(-q)}|_{v=0}\,,}
and similarly for the axial current. Consequently using \vpolamp\ we
get
\eqn\aadscft{\Pi_{\mu\nu}^V=P_{\mu\nu}\Pi _V(q^2)=\frac{\delta
^2S_V^{on-shell}}{\delta v_\mu^T(q)\delta v_\nu ^T(-q)}\,.}

Therefore correlation functions for vector and axial currents are
given by
\eqn\piv{\Pi _V(q^2)=-\frac{1}{2}\Lambda^3\dot v(q,\Lambda)\,,}
\eqn\pia{\Pi _A(q^2)=-\frac{1}{2}\Lambda^3\dot a(q,\Lambda)\,.}
Propagators for vector and axial-vector currents in the field theory
become the same if tachyon background vanishes. Non-vanishing
tachyon background breaks chiral symmetry, and therefore generally
speaking we have non-vanishing $S$ parameter, defined as
\eqn\sdef{S=-4\pi\frac{d}{dq^2}\left[\Pi _V(q^2)-\Pi
_A(q^2)\right]_{q^2=0}\,.}
With the help of holographic expressions \piv\ and \pia\ we obtain
\eqn\sdefhol{S=2\pi\Lambda^3\frac{d}{dq^2}(\dot v(q^2,\Lambda)-\dot
a(q^2,\Lambda))\,.}
%

The infrared behavior is specific for each particular tachyon
potential and we discuss it bellow. Now let us consider
near-boundary region. In the near-boundary region $r\gg 1$ we can
totally neglect tachyon field, which makes equations of motion for
vector and axial vector fields the same:
\eqn\vnbeq{\ddot v+\frac{3}{r}\dot v-\frac{q^2}{r^4}v=0\,,}
\eqn\avnbeq{\ddot a+\frac{3}{r}\dot a-\frac{q^2}{r^4}a=0\,.}
In practical computations one has to make sure that the last term in
\atreq\ is small, $T_0^2r^2/q^2\sim (q^2r^2)^{-1}\ll 1$ in
near-boundary region. This is important, because momentum $q$
competes in smallness with $1/r$ when one is computing S parameter.
Cutoff is supposed to be sent to infinity first, for each value of
momentum $q$. The solutions to these equations, normalized by
near-boundary condition \UVsource\ , are
\eqn\vbsnb{v=1-\frac{q^2}{2r^2}\log r+C_v\frac{1}{r^2}\,,}
\eqn\avbsnb{a=1-\frac{q^2}{2r^2}\log r+C_a\frac{1}{r^2}\,,}
where $C_v(q^2)$ and $C_a(q^2)$ define asymptotic near-boundary
behavior of the vector fields, have dimension two and go to finite
constants when $q^2=0$. Therefore substituting \vbsnb\ and \avbsnb\
into \sdefhol\ we find
\eqn\nbacd{S=4\pi\frac{d}{dq^2}(C_a-C_v)|_{q^2=0}\,.}
Notice that the $S$ parameter is expressed only via the coefficients
$C_{v,a}$, describing near-boundary behavior of vector and
axial-vector gauge fields, and does not depend on the cutoff
$\Lambda$.

Tachyon field describes chiral symmetry breaking at energy scale
given holographically by  $r\ll\mu$. In that region we have
essentially different dynamics of axial  vector and vector gauge
fields. In what follows we measure all dimensionful quantities in
units of dynamically generated scale $\mu$.

{\it Soft Wall}

Consider tachyon potential
\eqn\swtwo{V(T)=(1+(A-2)T^2)e^{-AT^2}\,,}
with $A>2$. Near the horizon in this potential tachyon field behaves
as $T_0(r)=1/r^{A/2}$. Correspondingly Lagrangian for vector field
fluctuation is
\eqn\swtvl{L_v=r^{3-\frac{A}{2}}e^{-\frac{A}{r^A}}\left(\dot
v^2+\frac{q^2A^2}{4}r^{-A-4}v^2\right)\,.}
It is useful to redefine
\eqn\vswtred{v=r^{\frac{A+2}{2}}e^{\frac{A}{2r^A}}\psi_v}
and consider Lagrangian for $\psi _v$
\eqn\swtpsiv{L_v=r^{\frac{10+A}{2}}\dot\psi_v^2+\frac{A^4}{4}r^{\frac{3(2-A)}{2}}\psi_v^2\,.}
Solution of the corresponding equation of motion is a linear
combination of Bessel functions
$I_{\pm\alpha}\left(\frac{A}{2r^A}\right)$ times a power of $r$, of
which the regular combination behaves as
\eqn\psivswt{\psi_v=r^{\frac{A}{4}-2}e^{-\frac{A}{2r^A}}\,.}
Correspondingly
\eqn\swtvnh{v=r^{\frac{3A}{4}-1}\,.}

Near horizon Lagrangian for axial field is
\eqn\swtaxl{L_a=r^{3-\frac{A}{2}}e^{-\frac{A}{r^A}}\left(\dot
a^2+\frac{A^2}{r^{2(A+1)}}a^2\right)}
It is convenient to make a redefinition
\eqn\swaxred{a=r^{\frac{A+2}{2}}e^{\frac{A}{2r^A}}\psi_a\,.}
The near-horizon Lagrangian for axial field is now
\eqn\nhredafl{L_a=r^{\frac{10+A}{2}}\dot\psi_a^2+\frac{A^2(A^2+4)}{4}r^{\frac{3(2-A)}{2}}\psi_a^2\,.}
Similarly to the case with vector field we choose the regular
solution, which is
\eqn\psiaswt{\psi_a=r^{\frac{A}{4}-2}\exp\left(-\frac{\sqrt{A^2+4}}{2r^A}\right)\,.}
Correspondingly near-horizon behavior of axial field is given by
\eqn\axfnhb{a=r^{\frac{3A}{4}-1}\exp\left(-\frac{\sqrt{A^2+4}-A}{2r^A}\right)\,.}
To summarize: we have the following near-horizon boundary
conditions:
\eqn\swtbct{T_0(r)=\frac{1}{r^{A/2}}\,,\quad
v(r)=r^{\frac{3A}{4}-1}\,,\quad
a(r)=r^{\frac{3A}{4}-1}\exp\left(-\frac{\sqrt{A^2+4}-A}{2r^A}\right)\,.}

We present results of numeric evaluations of the S parameter for
different values of $A$ in figure 3.

\ifig\loc{$S$ parameter in the soft wall potential \swtwo, depending
on the value of the parameter $A$.}
{\epsfxsize3.3in\epsfbox{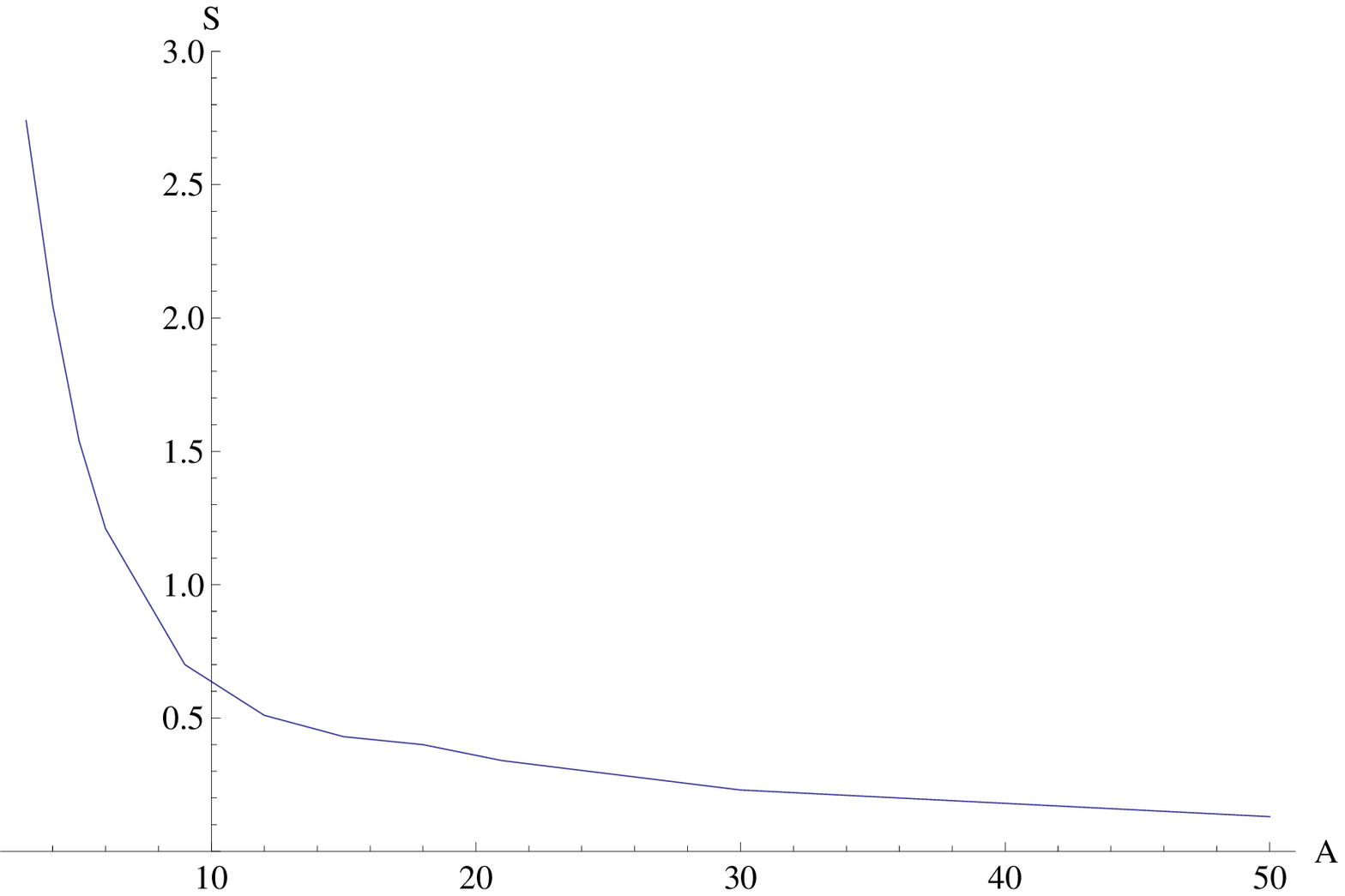}}

{\it Hard wall}

Consider hard wall tachyon potential
\eqn\hwpoten{V(T)=(\cos\, T)^4\,.}
The IR regime of the field theory corresponds to the near hard wall
region of AdS space, $r\simeq \mu$, where $\mu$ is the dynamically
generated scale. Let us measure all dimensional quantities in units
of $\mu$. Then the hard wall is located at $r=1$. When $r\simeq 1$,
the tachyon field behaves as
\eqn\nhwtb{T(r)\simeq\frac{\pi}{2}-c\sqrt{r-1}\,,\quad\quad
c=\sqrt{\frac{5}{2}}\,.}
Plugging \nhwtb\ into equations of motion for vector and
axial-vector gauge fields, \vtreq\ and \atreq, and considering the
region near $r=1$, we obtain
\eqn\vfnhwn{\ddot v+\frac{5}{2(r-1)}\dot v-\frac{5q^2}{8(r-1)}v=0}
\eqn\afnhwn{\ddot a+\frac{5}{2(r-1)}\dot
a-\frac{5(q^2+\pi^2)}{8(r-1)}a=0\,.}
The solutions are given by
\eqn\vanhwsol{v=\frac{c^v_1}{(r-1)^{1/2}}\left(1+\frac{d_1}{r-1}\right)+c_2^v+\OO
(\sqrt{r-1})\,,\quad
a=\frac{c^a_1}{(r-1)^{1/2}}\left(1+\frac{d_2}{r-1}\right)+c_2^a+\OO
(\sqrt{r-1})\,,}
where $d_1$ and $d_2$ stand for known functions of $q^2$. We require
momentum density $T^{0r}$ to vanish at $r=1$. The momentum density
is given by (to compute it perturb the background metric by small
$g_{0r}$ and keep only terms of the action which are linear in
$g_{0r}$)
\eqn\momdens{T^{0r}\sim\frac{1}{\sqrt{|g|}}\frac{\delta S}{\delta
g_{0r}}\simeq\frac{V(T)}{\sqrt{1+r^2\dot T^2}}\eta^{ij}\left[\dot
V_iF_{0j}^{(V)}+\dot A_iF_{0j}^{(A)}\right]\simeq (r-1)^{5/2}(v\dot
v+a\dot a)\,.}
We therefore choose the boundary conditions near the wall
\eqn\regvanhwsol{v=1\,,\quad\quad a=1}
Similarly as we have done in the soft wall case, we can now compute
the S parameter. Numerics give $S\simeq 2.6$.

\newsec{Lightest mesons}

\noindent In this Section we compute the sigma-mesons spectrum in
soft wall potential. Consider fluctuation of the tachyon field $\tau
(r,t)$ around the vacuum configuration $T_0(r)$. Expanding the TDBI
action
\eqn\tdbitfl{S=-\int d^4xdrV(T)r^3\left(1+r^2(\dot
T_0+\dot\tau)^2-\frac{1}{r^2}(\p_t\tau)^2\right)^{1/2}}
we arrive at the action for fluctuation field
\eqn\fltact{S=-\int d^3xd\omega
dr\left(G(r)\dot\tau^2+U(\tau)\tau^2\right)\,.}
Perform a Fourier transform
\eqn\tflftr{\tau (r,t)=\int\frac{d\omega}{2\pi}\tau
_\omega(r)e^{i\omega\tau}\,,}
where $\omega ^2=m^2$ is the squared mass of the tachyon excitation
mode. For the soft wall potential (we consider $A>2$ to get a
discrete spectrum of sigma-mesons, see \KutasovUQ\ for details)
\eqn\betatp{V(T)=(1+(A-2)T^2)e^{-AT^2}}
we obtain
\eqn\Grsw{\eqalign{G(r)&=\frac{e^{-AT_0^2}r^5(1+(A-2)T_0^2)}{2(1+r^2\dot
T_0^2)^{3/2}}\cr
  U(r)  & =\frac{\p}{\p
r}\left(\frac{e^{-AT_0^2}r^5T_0\dot
T_0(2+A(A-2)T_0^2)}{\sqrt{1+r^2\dot T_0^2}}\right)\cr
              &{+}e^{-AT_0^2}r^3(-2{+}AT_0^2(10{-}3A{+}2(A{-}2)AT_0^2))\sqrt{1{+}r^2\dot
T_0^2}{-}m^2\frac{e^{-AT_0^2}r(1{+}(A{-}2)T_0^2)}{2\sqrt{1{+}r^2\dot
T_0^2}}}}
Near the horizon $r=0$ the background tachyon field behaves as
$T_0=\frac{1}{r^{A/2}}$. Therefore the Lagrangian for fluctuating
field is
\eqn\fkLnh{L=\frac{4}{A^2}r^{\frac{10+A}{2}}e^{-A/r^A}\dot\tau^2-m^2r^{\frac{2-A}{2}}e^{-A/r^A}\tau^2\,.}
It is convenient to make a redefinition $\tau =e^{A/(2r^A)}\psi$ and
consider the field $\psi$ with the Lagrangian
\eqn\Lpsil{L_\psi=r^{\frac{10+A}{2}}\dot\psi^2+\frac{A^4}{4}r^{\frac{3(2-A)}{2}}\psi^2}
The solution of equation of motion for the field $\psi$ is a linear
combination of Bessel functions $I_{\pm\alpha}(A/(2r^A))$, times a
power of $r$. We choose the regular combination of Bessel functions,
which is
\eqn\besc{I_\alpha (A/(2r^A))-I_{-\alpha}(A/(2r^A))\simeq
r^{A/2}e^{-A/(2r^A)}\,.}

Corresponding near-horizon behavior of fluctuation tachyon field is
\eqn\tauflnhb{\tau (r)=r^{\frac{A}{4}-2}\,.}
We therefore impose the near-horizon conditions
\eqn\mtsmf{\tau(\epsilon)=1\,,\quad\tau'(\epsilon)=\left(\frac{A}{4}-2\right)\frac{1}{\epsilon}\,.}
We then integrate equation of motion for $\tau$ with these boundary
conditions up to the near-boundary region. We fit the result with
the expression
\eqn\taunbf{\tau(r)=\frac{1}{r^2}(c_1\log r+c_2)\,.}
The ratio $c_1/c_2$ must be equal to this ratio for the background
field $T_0$. This determines the discrete mass spectrum of tachyon
excitations.



We compute numerically the values $m_1^2$ and $m_2^2$ of the masses
of the first two excitations as a function of the parameter $A$ of
the tachyon potential \betatp. We plot the result of numerics in
figure 4.

\ifig\loc{The values of of the masses of the first two excitations
of the tachyon as a function of the parameter $A$ of the tachyon
potential \betatp. The $m_1^2$ (rescaled by a factor of ten) is
plotted in red and the $m_2^2$ is plotted in black.}
{\epsfxsize3.3in\epsfbox{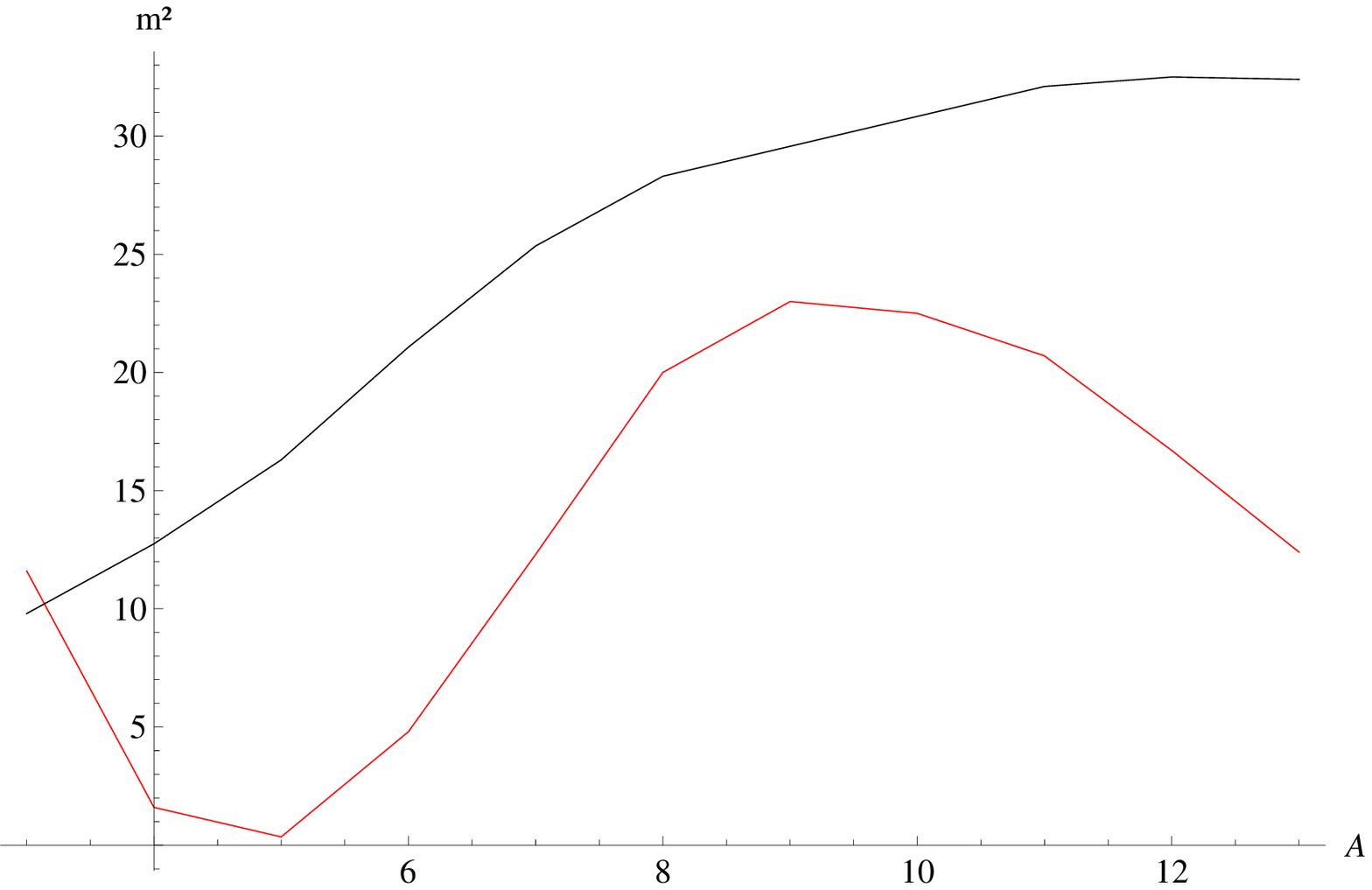}}

\newsec{Conclusion}

In this paper we have considered strongly coupled systems which
are described holographically by the tachyon DBI action in the AdS space-time.
These models are renormalizable: the UV cutoff can be taken to infinity
while the dynamically generated mass scale stays fixed.
We investigated the phase diagram of these models at
finite temperature and charge density.   For smaller values of
temperature and chemical potential the system resides in the
phase with broken conformal symmetry. This phase is separated by a phase transition line
from the phase with restored symmetry.
We observe that depending on the form of the tachyon potential, the order
of the phase transition may change, and hence one or more  critical point appears in
the diagram.
We have also used the TDBI action to describe
holographically dynamical electroweak symmetry breaking.
We have computed the S parameter using our holographic TDBI model
for generic soft-wall tachyon potential, and for  hard
wall tachyon potential.
The S-parameter takes generic positive values and does not appear to vanish in the
parameter space that we investigated.
We have also computed the masses of the lowest lying scalar mesons and observed that
even though there is no parametrically light scalar, the lightest meson can be made at least
an order of magnitude lighter than the next one.

\bigskip
\bigskip

\noindent {\bf Acknowledgement:}  We thank B. Galilo, D. Kutasov, J.
Maldacena and  D. Mateos for discussions. A.P. thanks Aspen Center
for Physics, where part of this work has been completed, for
hospitality. This work was supported in part by the NSF grant No.
1066293 and a VIDI innovative research grant from NWO.

\appendix{A}{Conformal phase transition and double-trace coupling running}

\noindent Consider a gauge field theory, coupled to matter fields with a
single-trace UV Lagrangian. When we go to lower energies,
integrating out higher momentum modes, we generally notice
\refs{\DymarskyUH\DymarskyNC-\PomoniDE}
 that effective Wilsonian Lagrangian contains double-trace
operators. We have to study the RG running of coupling constants for
double-trace operators if we want to study the fate of the theory at low energies.
Depending on the parameters defining the theory the
beta-functions for double-trace operators can exhibit essentially
different behavior; varying these parameters can lead to phase transitions
between different IR phases of the theory.
Here our focus will be on the particular type of these phase transitions,
called conformal phase transitions in \refs{\MiranskyPD}.
In this Section we review the field theory expectations for the physics
associated with conformal phase transitions (CPT).
We then use the technology of holographic Wilsonian RG to see how these
expectations are reproduced in a particular holographic model based on
the Tachyon DBI action in AdS space.

\subsec{Conformal phase transitions and Wilsonian RG}

\noindent Consider a gauge theory with $SU(N_c)$ gauge group,
coupled to $N_f$ massless Dirac fermions in the Veneziano limit,
where both $N_c$ and $N_f$ are taken to infinity, with the ratio
$x=N_f/N_c$ held fixed. It has a qualitatively different RG behavior
depending on the value of $x$. Let us look at the IR effective field
theory; three possible regimes can be  identified. When $x>11/2$ the
theory loses asymptotic freedom and is free in the IR; when
$x_c<x<11/2$, where $x_c\simeq 4$ (see \eg\ \JarvinenQE) is not
known precisely, the IR theory is in the interacting Coulomb phase.
This interval in $x$ , where the theory flows to a conformal fixed
point in the IR, is called "conformal window". However for $x$
smaller than $x_c$ the IR theory acquires a mass gap and chiral
symmetry is broken, due to the presence of chiral condensate.

The model studied in \PomoniDE\ is slightly different from the
example above, but exhibits similar behavior. The advantage is that
the beta function for the double-trace operator can be computed
exactly \PomoniDE. Suppose that we have some strongly interacting
theory, for which all single-trace operators have vanishing
beta-functions, e.g., orbifold theories \refs{\KachruYS,\LawrenceJA}
or non-supersymmetric deformations of $\NN=4$ super Yang-Mills
theory \LuninJY. To see whether the theory has conformal fixed
points we therefore have to study double-trace couplings
\refs{\DymarskyUH\DymarskyNC-\PomoniDE}. Denote by $\OO$ a
single-trace operator, and consider a double trace term in the
Lagrangian, $L_{dt}=f\OO^2$, where $f$ is a double-trace coupling
constant. In \PomoniDE\ (and, earlier, to the one-loop level in
\refs{\DymarskyUH,\DymarskyNC}) it has been shown that depending on
the parameters of the theory the beta function for $f$ either has a
real zero  (and then the theory flows to a conformal fixed point) or
it does not (and then the theory generates a mass gap).

We will observe a similar behavior in the holographic model based on
the tachyon DBI action in AdS space-time. First we  introduce a bulk
scalar field, dual to the field theory operator $\OO$. We choose it
to be the tachyon field $T$, described by the tachyon DBI action.
Now, we want to study renormalization of the corresponding
double-trace coupling $f$. We will use the holographic Wilsonian
renormalization as described in \refs{\FaulknerJY}. \foot{See also
\refs{\HeemskerkHK}.} The full AdS action is written as a sum of the
bulk action (in our case it is the tachyon DBI action), defined up
to the cutoff $\Lambda$, and the boundary action at $r=\Lambda$,
\eqn\AdSHRG{S[T]=\int _0^\Lambda drd^{d}x L_0[T]+\int
d^dxL_B[T]_{r=\Lambda}\,.}
To obtain holographically correlation functions that are invariant
under the RG flow, one has to require invariance of the action $S$ under
the change of $\Lambda$: this is a holographic implementation of
the Callan-Symanzik equation. The boundary term $S_B$ encodes all
degrees of freedom from the integrated out region $r>\Lambda$ of the
AdS space, and is written down as a sum of multi-trace operators
with corresponding coupling constants multiplying these operators.
Solving for $S_B$ the holographic RG (HRG) equation we determine running of the dual
field theory coupling constants.

Below we apply this method to the tachyon-DBI action in the AdS space and
find the RG behavior of the double-trace coupling $f$, depending on
the mass $m$ of the tachyon field. The non-vanishing tachyon field
in the bulk is a preferred state when $m^2<m_{BF}^2=-\frac{d^2}{4}$
\refs{\KutasovUQ}.
We conclude that $f$ exhibits a
walking behavior between the IR scale $\Lambda_{IR}$ and the UV
cutoff scale $\Lambda_{UV}$
which are related as
$\Lambda_{IR}=\Lambda
_{UV}\exp{\left(-\frac{\pi}{\sqrt{m_{BF}^2-m^2}}\right)}$.
Such a relation confirms that our holographic model exhibits a
conformal phase transition. This was also observed in \KutasovUQ,
where a similar relation between the UV cutoff and the physical
observables of the theory, such as e.g. meson masses, was
established.

Finally we remind the reader what happens as the tachyon mass squared
is lowered below the BF bound.
According to the AdS/CFT dictionary, the dimension of the operator $\OO$,
dual to the tachyon field $T$, is given by
\eqn\OOdim{\Delta_\pm=\frac{d}{2}\pm\sqrt{\frac{d^2}{4}+m^2}\,.}
The two  possible scaling
dimensions in \OOdim, $\Delta _-$ and $\Delta_+$ of the operator $\OO$ are
realized in the two conformal fixed points: the UV and IR respectively.
When we turn on a double-trace deformation $f\OO^2$ in the UV theory, the theory
flows to the IR conformal fixed point, where dimension of $\OO$
becomes equal to $\Delta_+$ \refs{\WittenUA}.
When the value of $m^2$ is lowered below $-d^2/4$,
the two fixed points merge and  then disappear, and the Miranski scaling emerges
\refs{\KaplanKR}.

\subsec{Double trace running from tachyon DBI}

\noindent Consider the tachyon-DBI bulk action for the tachyon field $T(r)$ of
the mass $m$, defined up to UV cutoff scale $r=\Lambda$ in
$AdS_{d+1}$:
\eqn\tdbiaoonshL{S_0=-\int _0^\Lambda dr\int
d^dxr^{d-1}V\sqrt{1+r^2\dot T^2}\,,}
where tachyon potential is expanded around $T=0$ as
\eqn\Tpotzdzt{V(T)=1+\frac{m^2T^2}{2}+\cdots\,,}
and we denote differentiation w.r.t. $r$ by dot. Suppose we
integrate out all degrees of freedom in the bulk which correspond to
$r>\Lambda$. Then we generate holographic Wilsonian effective action
\eqn\HWeffactd{S=S_0+S_B[T,\Lambda]\,,}
where $S_B$ is boundary term, which encodes integrated out degrees
of freedom.

Boundary condition at $r=\Lambda$ is given by
\eqn\hwbc{\Pi=\frac{\p S_B}{\p T}\,,}
where we have introduced momentum $\Pi$, canonically conjugate to
the tachyon field $T$:
\eqn\PiTconjd{\Pi=-\frac{\delta S_0}{\delta
T(r=\Lambda)}=\frac{\Lambda^{d+1}V\dot T}{\sqrt{1+\Lambda^2\dot
T^2}}\,.}
Using boundary condition \hwbc\ one may then express
\eqn\TderPi{\dot T=\frac{\p S_B/\p T}{\Lambda\sqrt{\Lambda
^{2d}V^2-(\p S_B/\p T)^2}}\,.}
If we denote $S_0=\int _{r_h}^\Lambda dr\int d^dxL_0$, then
holographic RG equation is
\eqn\HolRG{\frac{\p S_B}{\p\Lambda}+L_0(r=\Lambda)+\frac{\p S_B}{\p
T}\dot T(\Lambda)=0\,.}
With the help of \tdbiaoonshL\ and \TderPi\ this eventually acquires
the form
\eqn\Holrgs{\frac{\p S_B}{\p\Lambda}=\Lambda^{d-1}
V\sqrt{1-\frac{1}{\Lambda ^{2d}V^2}\left(\frac{\p S_B}{\p
T}\right)^2}\,.}
Action $S_B$ implicitly contains boundary metric factor
$\sqrt{-\det\, g_{b}}=\Lambda^d$. Let us make this factor explicit,
defining dimensionless boundary action $\SS$ as
\eqn\tsbdef{S_B=\Lambda ^d\SS\,.}
Let us also define new cutoff coordinate,
\eqn\epsdef{\epsilon=\log\frac{\Lambda}{\mu}\,,}
where $\mu$ is some constant, introduced for dimensional reasons.
HRG equation \Holrgs\ therefore gets rewritten as
\eqn\Holrgsn{\p_\epsilon\SS+d\SS=V\sqrt{1-\left(\frac{\p_T\SS}{V}\right)^2}\,.}

Let us expand the boundary action as
\eqn\SBexp{\SS=C(\epsilon)+J(\epsilon)T+\frac{1}{2}f(\epsilon)T^2\,.}
Plugging it into \Holrgsn\ and matching terms of the same order in
$T$, we obtain
\eqn\Ceq{\p_\epsilon C=\sqrt{1-J^2}-dC\,,}
\eqn\Jeq{\p_\epsilon J=-\frac{fJ}{\sqrt{1-J^2}}-dJ\,,}
\eqn\fEq{\p_\epsilon
f=\frac{m^2}{\sqrt{1-J^2}}-\frac{f^2}{(1-J^2)^{3/2}}-df}
We can solve these equations by putting $J\equiv 0$ and making $\bar
f=-(f+d/2)$ satisfy equation
\eqn\fneq{\p_\epsilon \bar f=\bar f^2-\frac{d^2}{4}-m^2\,.}
Let us denote $\kappa ^2=-\frac{d^2}{4}-m^2\equiv m_{BF}^2-m^2$,
then solution to \fneq\ may be written as
\eqn\fneqs{\bar f=\kappa\tan (\kappa\epsilon)\,.}
We conclude that double-trace coupling $\bar f$ exhibits a walking
behavior between UV scale
\eqn\LUVw{\Lambda_{UV}=\mu\exp\left(\frac{\pi}{2\kappa}\right)}
and IR scale
\eqn\LIRw{\Lambda_{IR}=\mu\exp\left(-\frac{\pi}{2\kappa}\right)}

\footatend\vfill\supereject\immediate\closeout\rfile\writestoppt
\baselineskip=14pt\centerline{{\bf References}}\bigskip{\frenchspacing%
\parindent=20pt\escapechar=` \input refs.tmp\vfill\eject}\nonfrenchspacing

\end